\documentclass{aa}  

\usepackage{graphicx}
\usepackage{txfonts}
\usepackage{lipsum}
\usepackage{subcaption}
                       
\usepackage{lscape}    
                       
\usepackage{placeins}  
                       
\begin{document}

\title{Parallel alignment of radio AGN jets and large cosmic web structures in the direction of the Shapley supercluster}
\titlerunning{Parallel radio AGN jets and large cosmic web structures in the direction of the SSC}

   \author{I. Baronchelli\inst{1,2}\corrauth{ivano.baronchelli@inaf.it}
        \and T. Venturi\inst{1,3}
        \and V. Casasola\inst{1}
        \and P. Merluzzi\inst{4}
        \and C. Burigana\inst{1,5}
        \and G. Busarello\inst{4}
        \and M. Bonato\inst{1,2}
        \and O. Smirnov\inst{1,3,6,7,8}
        \and T. Ronconi\inst{1,2}
        \and M. Massardi\inst{1,2}
        \and E. Liuzzo\inst{1,2}
        }

   \institute{$INAF-$Istituto di Radioastronomia, Via Piero Gobetti 101, I-40129, Bologna, Italy
     \and ALMA Regional Centre, Italian Node, Via Piero Gobetti 101, I-40129, Bologna, Italy
     \and Centre for Radio Astronomy Techniques and Technologies (RATT), Department of Physics and Electronics, Rhodes University, Makhanda, 6140, South Africa
     \and $INAF-$Osservatorio Astronomico di Capodimonte, Via Moiariello 16, I-80131 Napoli, Italy
     \and $INFN$, Sezione di Bologna, Via Irnerio 46, 40127 Bologna, Italy
     \and South African Radio Astronomy Observatory, Cape Town, 7925, South Africa
     \and Astrophysics, Department of Physics, University of Oxford, Keble Road, Oxford, OX1 3RH, UK
     \and Breakthrough Listen, Astrophysics, Department of Physics, University of Oxford, Keble Road, Oxford, OX1 3RH, UK
   }

   \date{Accepted: November 4th, 2026}

\abstract
  {}  
  {We investigated the potential large-scale alignment of radio active galactic nuclei (AGN) jets in the proximity of the Shapley Supercluster (SSC) area and its correlation with the identified cosmic web structures.} 
  {We analyzed 28 radio AGNs divided into two populations with comparable sky coverage north and south of a diffuse X-ray filamentary structure connecting clusters in the SSC core. We measured the average jet position angles ($PA$s) and combined them with spectroscopic redshifts to assess the persistence of the alignment along the line of sight.}
  {We detected a significant large-scale alignment of the radio AGN jets with the foreground X-ray filament (oriented at $PA \sim 115^{\circ}\pm 10^{\circ}$). Both the northern ($PA \sim 117^{\circ}$) and southern ($PA \sim 129^{\circ}$) groups align with this axis, yielding an overall statistical significance of $4.5^{6.1}_{3.8}\sigma$.  Although the projected alignment is highly significant, sparse spectroscopic coverage (seven sources spanning $0.048\lesssim z\lesssim 0.55$) suggests that this spatial coherence may extend beyond the SSC ($z\sim0.048$), hinting at a line-of-sight correlation across significantly larger cosmic scales.}
  {The observed alignment cannot be explained by localized dynamical processes alone. These findings may indicate a persistent influence of the large-scale structure of the cosmic web, or preferred directions of matter collapse, on supermassive black hole spin axes, highlighting a connection between AGN jet morphology and the underlying dark matter distribution. Such alignments represent an important systematic that must be accounted for in precision cosmology with surveys such as Euclid.}


 \maketitle
 \nolinenumbers
 
\section{Introduction}
\label{sec:intro}

In the standard Lambda -- Cold Dark Matter ($\Lambda$CDM) cosmological framework, the large-scale structure (LSS) of the Universe is organized into a network of nodes, filaments, and walls, known as the ``cosmic web'' \citep{Bond1996}. The emergence of this filamentary network was first predicted through the application of singularity theory to gravitational collapse \citep{Zeldovich1970, Arnold1982}, describing how matter density perturbations evolve into a hierarchical system of caustics (i.e., regions where the crossing of matter trajectories leads to formal singularities in density).

A fundamental question is whether the environmental influence of this LSS extends to the internal dynamics of galaxies and their central engines. According to the tidal torque theory \citep{Peebles1969, Doroshkevich1970}, the gravitational tidal fields of the matter distribution can impart a preferred orientation to the angular momentum (spin) of collapsing haloes. Numerical simulations \citep{Codis2012,Dubois2014,GaneshaiahVeena2018} and observational surveys \citep{Tudorache2025,Tempel2013a,Tempel2013b} have explored the potential link between filament geometry and the orientation of galaxies and their active galactic nuclei (AGNs). These works suggest that the spin axis of galaxies, and by extension, the spin of their supermassive black holes (SMBHs), tends to be aligned either parallel or perpendicular to host filaments, depending on the mass of the system and its accretion history. 

Despite these theoretical insights, the observational picture remains elusive. While localized or mass-dependent alignments have been reported, wide-field analyses frequently suggest that galaxy orientations are isotropic on larger scales \citep{Panwar2020}. Significant correlations between radio jet axes and cluster-scale environments have been detected in targeted studies \citep[e.g.,][]{Taylor2016}, yet recent wide-field surveys point toward a more stochastic distribution of orientations \citep{Jung2025}. Such findings imply that local dynamical processes or chaotic accretion episodes might effectively decouple the SMBH spin from the primordial cosmic web geometry.

The Shapley Supercluster (SSC) offers an exceptional laboratory to test all of these models. As one of the most massive and dynamically active structures in the local Universe ($z \approx 0.048$), the SSC hosts a high density of clusters connected by bridges of hot, X-ray emitting gas \citep{Bardelli1998, Bardelli2000, Rossetti2007}. In particular, ROSAT-RASS data have revealed a diffuse X-ray filamentary backbone linking the major cluster components in this region \citep{Kull1999}.

In this paper, we analyze the statistical alignment of radio AGNs in 2D projected proximity to this X-ray filament and discuss the possible theoretical implications of our findings. Understanding the physical origin and intrinsic large-scale coherence of galaxy alignments is important for precision cosmology, as these correlations can constitute a significant systematic effect in weak-lensing analyses with wide-field surveys such as Euclid \citep{EuclidCollab2026_A,EuclidCollab2026_B,EuclidCollab2026_C,EuclidCollab2026_D}. Radio-jet orientations provide a complementary probe of these correlations, tracing SMBH orientation and its connection to the surrounding cosmic web.

This paper is organized as follows. In Section~\ref{sec:data}, we describe the  dataset used in this study. Section~\ref{sec:Area_def_and_cosm_contxt} concerns the criteria adopted to define the survey area within the broader cosmological context. The selection criteria for the radio AGN sample are detailed in Section~\ref{sec:AGN_SAMPLE_CRITERIA}, while Section~\ref{sec:redshift} reports the available spectroscopic and photometric redshifts for the counterparts identified. In Section~\ref{sec:optical}, we examine the morphological properties of the optical hosts, when they are identified, and their implications. The statistical significance of the observed AGN alignments is computed in Section~\ref{sec:stats}. The possible physical interpretations of our observations are discussed in Section~\ref{sec:discussion}. Finally, we draw our main conclusion in Section~\ref{sec:conclusions}.

\section{Data}
\label{sec:data}
Our work is based on three MeerKAT observations performed in the L band (digitized frequency range 856-1712 MHz) with 4096 frequency channels across the band. Two of the pointings, centered on A\,3556 (RA$_{J2000}=13^h24^m06^s$, DEC$_{J2000}=-31^{\circ}40^{\prime}12^{\prime\prime}$) and A\,3562 (RA$_{J2000}=13^h33^m35^s$, DEC$_{J2000}=-31^{\circ}40^{\prime}30^{\prime\prime}$), are part of the OT1 observing cycle and were observed on July 7 and 9, 2019, respectively. The third pointing, centered on A\,3558 (RA$_{J2000}=13^h27^m54^s$, DEC$_{J2000}=-31^{\circ}29^{\prime}32^{\prime\prime}$), was observed during the MeerKAT scientific commissioning, on July 4, 2018.
The final total intensity images  used in this work have an average rms of $\sim 6$ $\mu$Jy~beam$^{-1}$ throughout the field of view - characterized by a primary beam (PB) with full width half maximum $FWHM \sim 1^{\circ}$ - and angular resolutions of $\sim 7^{\prime\prime}\times6^{\prime\prime}$. Further details on the data analysis and the parameters of the final images are reported in \cite[][V22]{Venturi2022}. The pointing on A\,3556 (not reported in V22) has an angular resolution of $7.02^{\prime\prime}\times6.70^{\prime\prime}$ in position angle ($PA$) $143^{\circ}$.
For the other two pointings, the resolution is $7.02^{\prime\prime}\times5.9^{\prime\prime}$ at $PA\ 2^{\circ}$ (A\,3558) and $7.0^{\prime\prime}\times6.4^{\prime\prime}$ at $PA\ 152^{\circ}$ (A\,3562). Crucially, the nearly circular beams do not introduce a preferred jet orientation. Our extended sources have $PA$s primarily determined by the alignment of multiple resolved components (e.g., cores and terminal lobes), rather than by the morphology of individual detections. Thus, any residual instrumental effect is limited to the pixel grid; given the homogeneous pixel scale along both axes ($1^{\prime\prime}$/pixel), we do not expect a pixel-induced orientational bias. The source extraction process is detailed in Appendix \ref{SECT:AppendixA}.

\section{Area definition and cosmic context}
\label{sec:Area_def_and_cosm_contxt}

To assess the potential alignment of radio sources with the large-scale X-ray structure, we defined two study areas symmetrically arranged around the projected center of mass (CoM) of the SSC. We defined our selection geometry with respect to the X-ray filament connecting the clusters Abell 3558 (Shapley 8; RA $13^h 27^m 54^s$, Dec $-31^{\circ} 29' 32''$, hereafter A\,3558) and Abell 3562 (RA $13^h 33^m 35^s$, Dec $-31^{\circ} 40' 21''$, hereafter A\,3562), the two most massive clusters in the field. This region was identified as a major bridge of hot X-ray-emitting gas in the \textit{ROSAT} All-Sky Survey \citep{Kull1999} and subsequently characterized by \textit{XMM-Newton} observations \citep{Rossetti2007, Merluzzi2016}. While the SSC core lies at $z\sim0.048$ \citep{Proust2006a}, the survey area extends to higher-redshift background systems, allowing us to probe the cosmic web across multiple epochs.

\subsection{Cluster environment and redshift distribution}
\label{sec:cluster_env_and_z_distr}

The dynamical landscape of our survey region is characterized by a pronounced line of sight (LoS) stratification beyond the SSC core ($z \approx 0.048$), which represents the dominant mass concentration in the local Universe.
  In particular, this volume contains several distinct background overdensities, including a peak at $z \approx 0.08$ \citep[][see figure 5 and figure 1 therein]{Quintana2000, Proust2006a}, possible superclusters at $z \approx 0.11-0.14$ \citep[see section 5.2 therein]{Quintana2000}, and additional galaxy clusters at $z \approx 0.17$ \citep{Higuchi2020}. Overdensity peaks are also found in \cite{Merluzzi2015} and \cite{Haines2018} between $z \approx 0.07$ and $z \approx 0.25$.

Table~\ref{tbl:clusters}, adapted and extended from \cite{Merluzzi2015}, summarizes the properties of clusters in the foreground and background of the SSC area. Most background systems are broad galaxy overdensities rather than virialized clusters and thus lack well-defined dynamical mass estimates.

As discussed in \cite{Proust2006a,Proust2006b}, the SSC mass contributes significantly to the local velocity field. However, its inferred mass alone cannot fully account for the observed peculiar velocity of the Local Group relative to the cosmic microwawe background frame \citep{Hoffman2001}. Recent dynamical reconstructions show that the velocity field does not converge onto a single dominant attractor, but instead reflects contributions from progressively larger scales extending beyond the SSC region \citep{Stiskalek2026}. In this context, the SSC can be interpreted as an intermediate node within a larger-scale anisotropic structure rather than a terminal attractor. Instead of requiring a single massive, virialized system beyond Shapley, this scenario suggests a coherent matter distribution extending to higher redshifts, capable of contributing to the large-scale gravitational field and consistent with the non-convergent nature of the local velocity flow.

\begin{table*}[!ht]
\caption{\label{tbl:clusters}Properties of the foreground and background clusters and LSS intersecting the SSC area. }
\centering
\small
\begin{tabular}{lcccccc}
\hline
\hline
Cluster / Structure & RA J2000 & DEC J2000 & redshift ($z$) & richness\textsuperscript{a} & $R$ [Mpc]\textsuperscript{f} & Mass [$10^{14} M_{\odot}$]\textsuperscript{b} \\
\hline
\textbf{Local SSC complex} & & & & & & \\
\hline
AS 0724 & 13 13 08.6 & -32 59 38\textsuperscript{c} & $0.04958 \pm 0.00052$\textsuperscript{d} & 0 & --- & 0.88\textsuperscript{e} \\
AS 0726 & 13 15 11.7 & -33 38 52\textsuperscript{d} & $0.04967 \pm 0.00046$\textsuperscript{d} & 0 & 0.96 & 0.46\textsuperscript{f} \\
A 3552  & 13 19 00.7 & -31 51 04\textsuperscript{c} & $0.04921 \pm 0.00040$\textsuperscript{d} & 1 & --- & 0.36\textsuperscript{e} \\
A 3554  & 13 19 27.6 & -33 29 49\textsuperscript{c} & $0.04814 \pm 0.00031$\textsuperscript{d} & 1 & 0.80 & 0.58\textsuperscript{f} \\
A 3556  & 13 24 00.2 & -31 39 22\textsuperscript{c} & $0.04789 \pm 0.00025$\textsuperscript{g} & 0 & 0.98 & 1.7\textsuperscript{f} \\
A 3558  & 13 28 02.6 & -31 29 35\textsuperscript{c} & $0.04804 \pm 0.00020$\textsuperscript{g} & 4 & 1.16 & 13.0\textsuperscript{g} \\
A 3559  & 13 29 53.1 & -29 30 22\textsuperscript{h} & $0.04713 \pm 0.00019$\textsuperscript{d} & 3 & 0.31 & 0.20\textsuperscript{f} \\
A 3560  & 13 32 22.0 & -33 05 24\textsuperscript{i} & $0.04854 \pm 0.00035$\textsuperscript{d} & 3 & 1.33 & 3.4\textsuperscript{f} \\
A 3562  & 13 33 47.0 & -31 40 37\textsuperscript{c} & $0.04822 \pm 0.00064$\textsuperscript{d} & 2 & 0.89 & $3.9 \pm 0.4$\textsuperscript{j} \\
SC 1327-312 & 13 29 45.4 & -31 36 12\textsuperscript{k} & $0.04951^{+0.00035}_{-0.00070}$\textsuperscript{g} & --- & 1.30 & 0.30\textsuperscript{f} \\
SC 1329-313 (a) & 13 31 36.0 & -31 48 45\textsuperscript{k} & $0.04933^{+0.00038}_{-0.00022}$\textsuperscript{g} & --- & 1.14 & 0.37\textsuperscript{f} \\
SC 1329-313 (b) & ---   &  ---  & $0.04452^{+0.00023}_{-0.00028}$\textsuperscript{g} & --- &  ---  &  ---  \\
\hline
\textbf{Background LSS} & & & & & & \\
\hline
Overdensity peak & --- & --- & $\approx 0.08$\textsuperscript{l} & --- & --- & --- \\
Continuous clusters distribution & --- & --- & $\approx 0.08 - 0.13$\textsuperscript{l} & --- & --- & --- \\
Possible superclusters & --- & --- & $\approx 0.011, 0.14$\textsuperscript{m} & --- & --- & --- \\
Multiple overdensity peaks & --- & --- & $\approx 0.07 - 0.25$\textsuperscript{n} & --- & --- & --- \\
Multiple overdensity peaks & --- & --- & $\approx 0.096 - 0.195$\textsuperscript{o} & --- & --- & --- \\
Cluster & --- & --- & 0.17\textsuperscript{p} & --- & --- & --- \\
\hline
\end{tabular}
\tablefoot{ \\
\tablefoottext{a} {\cite{Abell1989}; }
\tablefoottext{b} {All the masses are dynamically derived except for that of A 3562 (X-ray observations);} 
\tablefoottext{c} {\cite{DeFilippis2005};} 
\tablefoottext{d} {\cite{Proust2006a}, the redshift value reported refers to the cosmic microwawe background frame;} 
\tablefoottext{e} {\cite{Reisenegger2000}, cluster mass within a radius enclosing 500 times the critical density;} 
\tablefoottext{f} {\cite{Ragone2006}, estimates of virial radius and virial mass;} 
\tablefoottext{g} {\cite{Bardelli1998}, the mass value has been transformed into adopted cosmology;} 
\tablefoottext{h} {\cite{David1999};} 
\tablefoottext{i} {\cite{Bardelli2002};} 
\tablefoottext{j} {\cite{Ettori2000}, the mass value has been transformed into adopted cosmology;} 
\tablefoottext{k} {\cite{Breen1994};} 
\tablefoottext{l} {\cite{Quintana2000}, \cite{Proust2006a};} 
\tablefoottext{m} {\cite{Quintana2000}; } 
\tablefoottext{n} {\cite{Merluzzi2015};} 
\tablefoottext{o} {\cite{Haines2018};} 
\tablefoottext{p} {\cite{Higuchi2020}.}
}
\end{table*}

The SSC, comprising clusters such as A3558, A3560, and A3562, is the primary foreground anchor of our survey. The X-ray bridge connecting these nodes traces matter infall and gas heating, providing a dynamically motivated reference for our geometric selection. In the same sky area, several large-scale structures are detected beyond the SSC core ($0.07 \lesssim z \lesssim 0.25$). Thus, our radio survey samples the underlying cosmic web rather than being confined to local cluster-scale phenomena.

\subsection{Selection geometry}

\begin{figure*}[!ht]
\sidecaption
\includegraphics[width=12cm]{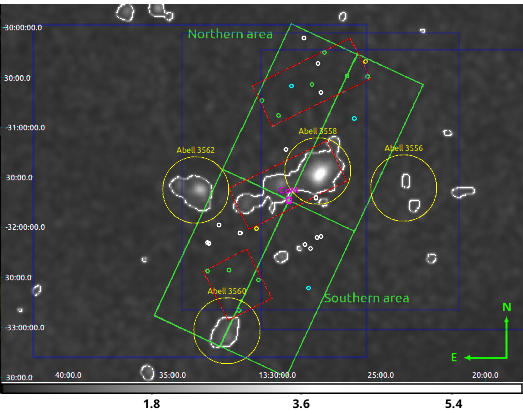}
\caption{Analysis regions and large-scale X-ray environment of the SSC. The image approximately covers the MeerKAT survey footprint. The ROSAT-RASS (0.1--2.4 keV) map has been smoothed with a Gaussian kernel ($\sigma = 2.5$ pixels, with a pixel scale of  $45^{\prime\prime}\mathrm{pixel}^{-1}$) to highlight the filament connecting the primary clusters (large yellow circles). The projected COM of A3556, A3558, A3560, and A3562 is marked by a magenta circle. Green boxes mark the northern and southern analysis regions, designed to sample the ``infall corridor'' connecting A3560 to the projected CoM of the SSC core. Green circles mark PA-coherent sources in the ``core'' samples (six out of six in the north; four out of five in the south). Red rectangles highlight their spatial distribution. White circles mark other sources in the ``extended'' samples. Cyan circles mark sources with complex morphologies (IDs 35348 and 35265 in the north; 34624 in the south). In the north, a small yellow circle marks source 35380, whose morphology is unclear; in the south, another small yellow circle marks source 53742, which meets the core-sample criteria but has a $PA$ roughly perpendicular to those of the other core sources.}
\label{img:Analysis_Regions}
\end{figure*}

Given the complex dynamical state of the region, characterized by a major merger event involving A3558 and A3562 \citep[e.g.,][]{Bardelli1998, Rossetti2007}, we adopted a 2D selection geometry (Figure~\ref{img:Analysis_Regions}) based on the mass-weighted projected CoM of the SSC core. We included the masses of A3558, A3562, A3560, and A3556, while neglecting smaller systems with masses $M < 10^{14} \, M_{\odot}$. Using the aggregate mass estimates (see Table~\ref{tbl:clusters}), we obtained the CoM at $\alpha_{J2000} = 13^{\text{h}} 29^{\text{m}} 25.0^{\text{s}}, \delta_{J2000} = -31^{\circ} 47' 06''$.

We defined a principal dynamical axis connecting the projected CoM to the southern cluster A3560, which acts as a secondary mass concentration and marks the direction of matter accretion toward the central filamentary complex. This axis defines the orientation of our analysis area, which extends symmetrically by $\sim 1.6^{\circ}$ north and south, for a total of $\sim 3.2^{\circ}$, reaching the survey footprint boundaries. The area spans $1.5^{\circ}$ along the orthogonal axis, dividing it into two symmetric $1.6^{\circ} \times 1.5^{\circ}$ domains:

\begin{itemize}
\item{The southern region encompasses a rectangular  box extending from the A3558/A3562 projected CoM toward A3560. The area is delimited by the vertices $(13^h 32^m 30^s, -31^{\circ} 28' 20'')$, $(13^h 26^m 19^s, -32^{\circ} 05' 26'')$, $(13^h 29^m 33^s, -33^{\circ} 33' 20'')$, and $(13^h 35^m 49^s, -32^{\circ} 55' 39'')$.}

\item{The northern region comprises a symmetric rectangular box extending northward. The area is delimited by the vertices $(\alpha, \delta)_{J2000}$: $(13^h 29^m 17^s, -30^{\circ} 00' 42'')$, $(13^h 23^m 09^s, -30^{\circ} 37' 14'')$, $(13^h 26^m 18^s, -32^{\circ} 05' 26'')$, and $(13^h 32^m 30^s, -31^{\circ} 28' 20'')$.}
\end{itemize}

The analysis region defined above roughly corresponds to the site of long cosmic filaments. Southward, these filaments connect A3560 to the central A3562-A3558-A3556 complex. Symmetrically, A3559 lies northward of the complex - outside the MeerKAT footprint - at a distance comparable to the southern separation between A3558 and A3560 \citep{Merluzzi2015, Haines2018, Quintana2020, Baier-Soto2025}.

Figure~\ref{img:Analysis_Regions} shows the analysis domains within the SSC core's dynamical environment. The background shows a smoothed ROSAT-RASS broadband (0.1--2.4 keV) X-ray map. While the horizontal filamentary bridge between A3562 and A3558 is evident, our selection is driven by the vertical dynamical axis. This axis connects the CoM of the A3558/A3562 complex (magenta circle) to the southern cluster A3560, marking the direction of large-scale matter infall toward the filamentary backbone.
The green rectangles delineate our study areas along this ``infall corridor,'' ideally sampling radio AGNs as they interact with the cosmic web.

\begin{figure}[!ht]
\centering
\resizebox{\hsize}{!}{\includegraphics{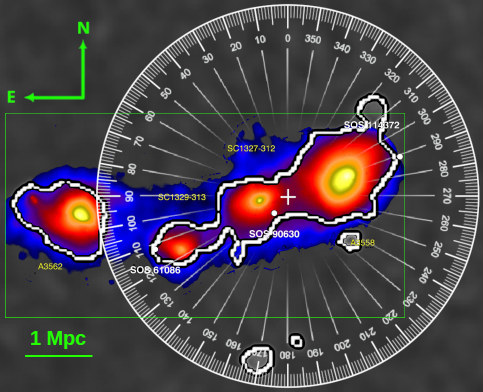}}
\caption{Filament connecting A3562 and A3558 as observed with \emph{XMM-Newton} \citep[0.5--2.5 keV, adapted from Fig.~1 of][]{Merluzzi2016}. The green box outlines the original image boundaries. The background map and white contours show the same ROSAT-RASS observations (0.1--2.4 keV) as in Figure~\ref{img:Analysis_Regions}. A goniometer indicates the $PA$ of the western sector of the filament ($PA\sim115^{\circ}\pm10^{\circ}$).}
\label{img:filament_alignement}
\end{figure}

We tested the robustness of our spatial selection by recalculating the projected CoM using alternative masses inferred from velocity dispersions ($\sigma_{v}$) in Table~4 of \citet{Haines2018}. The resulting CoM, at $\alpha_{\mathrm{J2000}} = 13^{\text{h}} 29^{\text{m}} 50.7^{\text{s}}, \delta_{\mathrm{J2000}} = -31^{\circ} 59' 23.5''$, lies $\sim13.5'$ south of our fiducial CoM but has a perpendicular offset of only $\sim45''$ from the primary CoM--A3560 axis. Adopting this position therefore leaves the orientation and boundaries of our survey areas largely unchanged, as these are primarily constrained by the MeerKAT footprint's north--south limits.
Figure~\ref{img:filament_alignement} shows an \textit{XMM-Newton} close-up of the central region in Figure~\ref{img:Analysis_Regions}, with a goniometer indicating the $PA$ of the western segment of the filament ($PA\sim115^{\circ}\pm10^{\circ}$).

\section{AGN sample selection criteria}
\label{sec:AGN_SAMPLE_CRITERIA}

Within the analysis area, we identified two homogeneous ``core'' samples of extended radio sources, one in each region. The most extended sources occupy two distinct, well-defined subregions (red rectangles in Figure~\ref{img:Analysis_Regions}), each with an apparent size comparable to that of the filament.
Additionally, we anticipate that these two samples show slightly different typical $PA$ ($\Delta \langle PA \rangle \sim20^{\circ}$) and apparent size.
For these reasons, we initially analyze the northern and southern samples separately.
The selection of the ``core'' samples is based on the following constraints:

\begin{itemize}
    \item {Significance: A lower flux density threshold of $5\sigma$ was imposed on at least one component of each radio source (either the core or the jets) to ensure robust detection against the local background noise.}
    \item {Morphological coherence: We selected sources exhibiting clear and consistent jet morphology. Specifically, we included objects where the geometric center and the jet axis could be unambiguously identified visually. Sources displaying perturbed, bent, or otherwise incoherent geometries were excluded from the  ``core'' sample and analyzed separately.}
    \item {Angular scale: We targeted the largest structures within each field. Specifically, for each region, we defined a selection threshold $L^\mathrm{th}$ corresponding to $80\%$ of the average extension of the five largest sources ($L^{5}$) fulfilling the previous two criteria ($L^\mathrm{th}=0.8L^{5}$). This rule translates into angular scales exceeding 57'' and 46'' in the northern and southern regions, respectively.}
\end{itemize}

Applying these criteria, we selected six and five AGNs in the northern and southern sectors, respectively. To assess potential biases from the initial ``core'' selection and test the robustness of our results against variations in the sample definitions, we relaxed these criteria defining broader ``extended'' samples. These included the 14 largest sources in each sector meeting the significance criterion, regardless of morphological coherence (including one-sided jets and perturbed, bent, or otherwise incoherent morphologies). The extended selection resulted in minimum apparent sizes of $37''$ and $24''$ in the northern and southern sectors, respectively. Complex morphologies prevented $PA$ measurements for one source in each sector, leaving 13 usable sources per extended sample.

The distribution of the radio sources relevant for our analysis within the gravitational landscape is color-coded in Figure~\ref{img:Analysis_Regions}.
Notably, the distribution of the northern ``core'' sources (six out of six sources with $PA$s parallel to the filamentary axis) appears to replicate the filamentary geometry. The distortions observed in morphologically complex sources (belonging to the ``extended'' samples) may reflect dynamical shear or ram pressure from the surrounding medium. Only sources 35348 and 34624 were excluded from the statistical analysis because their $PA$s could not be determined.

\subsection{Northern region sample}
\label{sec:EXT_NORTH_SAMPLE}

\begin{figure*}[!ht]
\centering
\includegraphics[width=17cm]{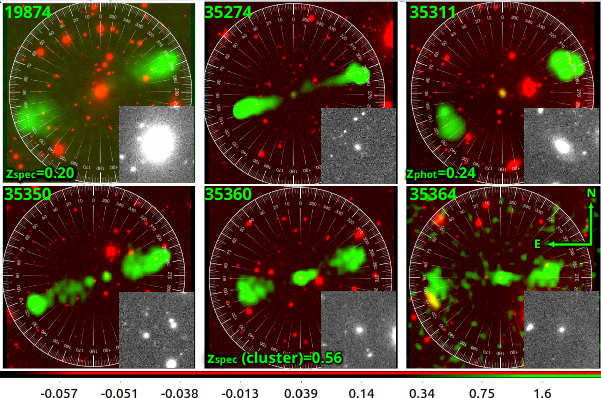}
\caption{Composite of the individual 170''×170'' radio stamps of the six northern region sources selected for the stacking procedure (IDs: 19874, 35274, 35311, 35350, 35360, and 35364), in green, with the VST image ($r_{SDSS}$) of the same region, in red. Each radio image has been normalized to the 98$^{\mathrm{th}}$ percentile of its surface-brightness distribution to highlight the morphological structure of the jets and lobes independently of the absolute flux density. In all panels, north is up and east is to the left, as indicated by the orientation arrows. The colorbar represents the normalized intensity. To help appreciate the similar orientation, a goniometer image has been overlaid to each stamp. In each panel, the redshift is reported when known, prioritizing the spectroscopic measurement if available or the photometric one if not. If neither the spectroscopic nor the photometric redshsift is available, the spectroscopic redshift of the cluster to which the source is supposed to belong is reported instead. In each panel, in the bottom-right inset, a magnified version of the central region of VST $r_{SDSS}$ image (30'' by 30'') is also shown. A clear optical counterpart is always visible, with the only exception being ID 35350, where multiple possible candidates are detected. The major axes are approximately orthogonal to the orientation of the jets in 35311, while a $\sim45^{\circ}$ angle is visible for 19874. No clear morphology is discernible for the remaining counterparts. }
\label{img:stacked_north_sample_all}
\end{figure*}

In the northern sector, the five largest sources meeting the morphological and significance criteria have a mean apparent size of $L^{5}_{\mathrm{north}}=71.6''$. We therefore set the ``core'' sample size threshold to $L^{\mathrm{th}}_{\mathrm{north}}=0.8L^{5}_{\mathrm{north}}=57''$.
Applying these criteria, we selected six sources (IDs: 19874, 35274, 35311, 35350, 35360, and 35364) for the ``core'' sample. Their primary axes show a remarkably coherent orientation, parallel to the western segment of the filament connecting A3558 and A3562. Figure~\ref{img:stacked_north_sample_all} shows the $170''\times170''$ stamps of these sources.

In Appendix ~\ref{SECT:AppendixB}, we show (Figures~\ref{img:not_stacked_north_sample_all} and \ref{img:not_stacked_north_sample_B}) and describe the sources that, although failing to meet the formal criteria for the ``core'' sample, were included in the ``extended'' sample.
Among these, IDs 35265 and 35348, with angular extensions meeting the size selection criterion, are characterized by asymmetric or disturbed profiles; their complex structures may serve as high-resolution, localized probes of the filament's dynamical environment.

We stacked the six northern region sources of the ``core'' sample. For each system, we extracted a $170 \times 170$ pixel radio stamp (1 arcsec pixel$^{-1}$), corresponding to a projected area of $170'' \times 170''$, centered on the midpoint of the radio complex.

Crucially, no rotational alignment or size-scaling was applied prior to stacking. The jets of the selected sources are already coherently oriented on the sky, parallel to the large-scale filamentary structure traced by the X-ray emission between A3558 and A3562. The stacking procedure is indeed intended to test and enhance this intrinsic alignment, rather than to impose it artificially. Interestingly, although our criteria simply targeted the most extended sources (with consistent morphology), their apparent sizes are remarkably homogeneous, ranging between 62'' and 81'' without extreme outliers.

The final stacked image was obtained as the pixel-by-pixel mean of the six normalized stamps. To prevent brighter objects from dominating the stack, each stamp was normalized to the 98$^{\mathrm{th}}$ percentile of its surface-brightness distribution. This provides a robust proxy for characteristic high-brightness emission while reducing sensitivity to extreme pixel values and local noise spikes. Thus, all sources contribute comparably to the composite image, preserving relative morphological structure while suppressing absolute flux differences. We verified that total-flux normalization yields a morphologically consistent stack, confirming that the detected alignment does not depend on the normalization scheme.

\begin{figure}[!ht]
\centering
\resizebox{\hsize}{!}{\includegraphics{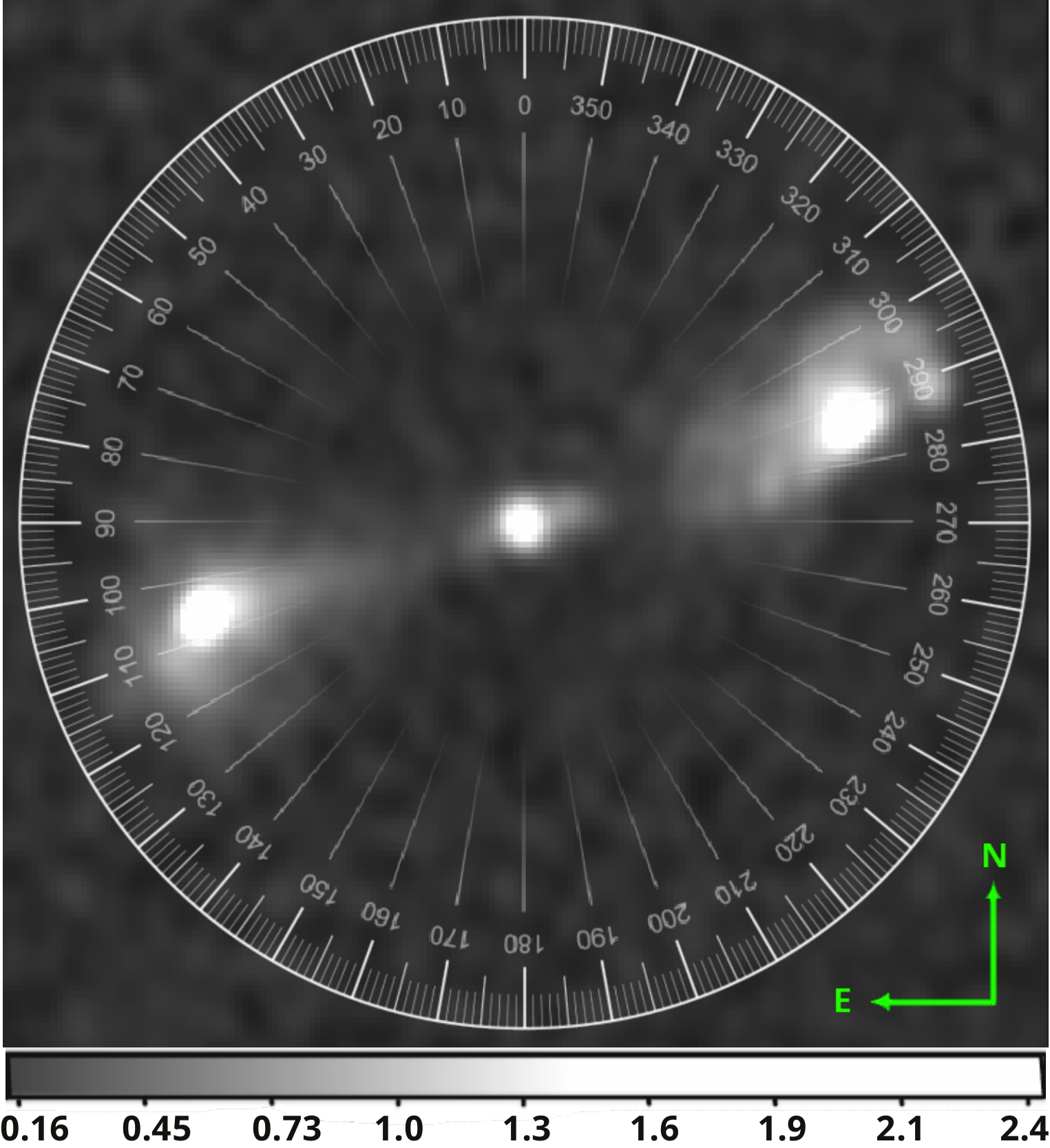}}
\caption{Stacked radio image of the six northern region AGNs in the ``core'' sample, obtained as explained in Section~\ref{sec:EXT_NORTH_SAMPLE}. North is up and east is left.}
\label{img:stacked_north_sample}
\end{figure}

The resulting composite, shown in Figure~\ref{img:stacked_north_sample}, enhances an emission component that is spatially coherent across the sample. The stack reveals a central peak corresponding to the average core position and an elongated structure extending along the common jet axis. The persistence of this elongated emission in the absence of any imposed rotational alignment demonstrates that the preferred orientation is an intrinsic property of the sample. Visual inspection suggests an apparent $PA$ in the range $105^{\circ}-110^{\circ}$. 

\subsection{Southern region sample}
\label{sec:EXT_SOUTH_SAMPLE}

\begin{figure*}[!ht]
\sidecaption
\includegraphics[width=12cm]{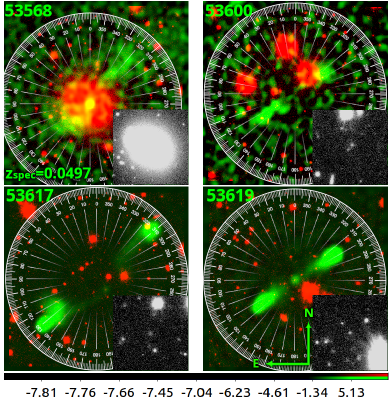}
\caption{Individual radio stamps of the four sources selected for the southern region stacking (IDs: 53568, 53600, 53617, and 53619, in green). The VST image ($r_{SDSS}$) of the same region is overlaid (in red). All visual conventions are identical to those defined in Figure~\ref{img:stacked_north_sample_all}. A clear optical detection is always visible, with the exception of ID 53600, where no clear counterpart is detected. In 53568 the major axis is approximately orthogonal to the orientation of the jets, whereas no clear morphology is discernible for the remaining counterparts.}
\label{img:stacked_south_sample_all}
\end{figure*}

In the southern region, we identified five radio sources meeting all ``core'' selection criteria. Four are closely aligned (IDs: 53568, 53600, 53617, and 53619), whereas ID 53742 lies approximately perpendicular to their common direction. The average apparent size of these five sources, which are the largest meeting the morphological and significance criteria, is $L^{5}_\mathrm{south}=57.8''$, smaller than in the northern region.
We therefore adopt a lower angular size threshold ($L^\mathrm{th}_\mathrm{south}=0.8L^{5}_\mathrm{south}=46''$, versus $57''$ for the northern sample), keeping the remaining selection criteria unchanged. Figure~\ref{img:stacked_south_sample_all} shows the individual $170'' \times 170''$ radio stamps of the aligned ``core'' sources.

As in the northern sample, these sources exhibit a remarkably coherent orientation, with their primary axes aligned with the western portion of the filamentary backbone connecting A3558 and A3562. Their apparent sizes cluster between $47''$ and $55''$, except for ID 53617 ($80''$). This similarity partly reflects our size-selection criterion.

Also in the southern area, the remaining largest sources failing to meet the formal ``core'' criteria, but representing significant ``borderline'' cases, are included in the ``extended'' sample.
These objects were initially excluded either due to their smaller angular extent or because of morphological complexities and geometric irregularities (compared to the simple centroid-jets configuration). These sources are described in Appendix~\ref{SECT:AppendixB} and shown in Figures~\ref{img:NOT_stacked_south_sample_A} and \ref{img:NOT_stacked_south_sample_B}. Source ID 53742 (size $\sim 55''$), despite belonging to the ``core'' sample, is shown in this group due to its complete misalignment with the filament.

The stacking analysis for the southern region followed the methodology detailed in Section~\ref{sec:EXT_NORTH_SAMPLE}. For this region, we considered only the four aligned sources in the ``core'' sample, excluding source ID 53742 due to its orthogonal $PA$.

\begin{figure}[!ht]
\centering
\resizebox{\hsize}{!}{\includegraphics{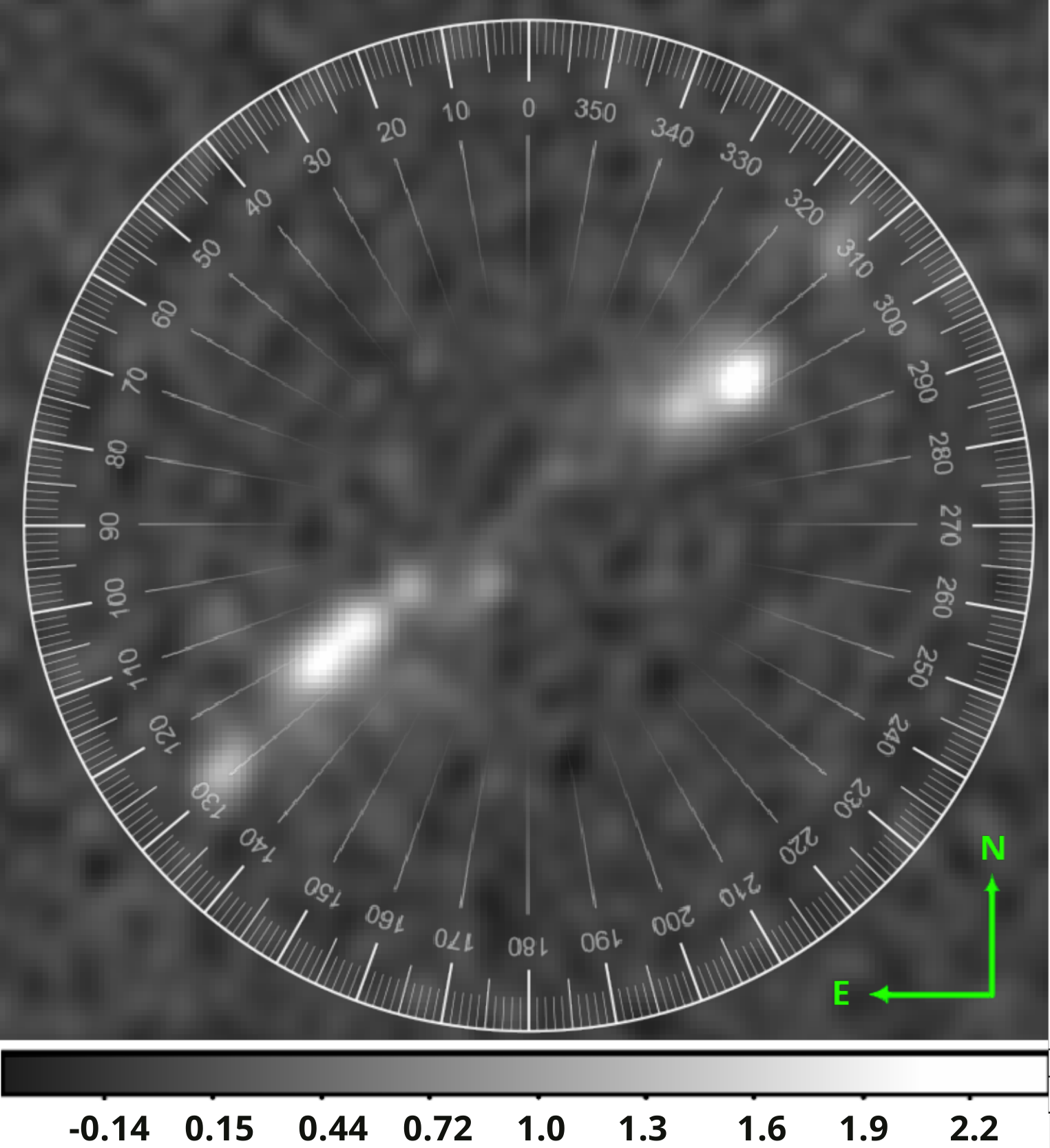}}
\caption{ Stacked radio image of the four aligned AGNs of the southern ``core'' sample obtained as described in Section~\ref{sec:EXT_SOUTH_SAMPLE}. North is up and east is left.}
\label{img:stacked_south_sample}
\end{figure}

As with the northern sample, we extracted $170 \times 170$ pixel stamps centered on the geometric center of each radio complex deliberately avoiding rotational alignment to preserve the original orientation of the jets as observed on the sky.
To ensure a balanced contribution to the final composite, each stamp was normalized to the 98th percentile of its surface-brightness distribution. The final stack (Figure~\ref{img:stacked_south_sample}) was obtained by pixel-by-pixel averaging these normalized images.

The final stack confirms a strikingly coherent elongated structure despite the smaller sample (four sources), with a larger $PA$ ($125^{\circ}\pm 5^{\circ}$) than in the northern region. Unlike the northern stack (Fig.~\ref{img:stacked_north_sample}), which shows a prominent central peak, the southern composite lacks a well-defined central core.

This behavior is primarily due to the fact that, despite the symmetric selection criteria applied to both regions, the southern sources exhibit a non-negligible intrinsic asymmetry compared to their northern counterparts. In this southern subsample, the jets are not symmetrically distributed around the central engine.
In ID 53568 and the slightly warped ID 53617, the jet-lobe structure is significantly offset or uneven, so the geometric center of the emission, adopted as the stacking reference point, does not precisely coincide with the core position. Furthermore, for ID 53600, no clear central emitter is identifiable. Consequently, stacking smears out the core positions, whereas the global orientation of the emission remains statistically robust.

\section{Redshift analysis and environmental context}
\label{sec:redshift}

To investigate the three-dimensional distribution of our radio sample and identify potential host galaxies, we cross-matched the radio centroids with available spectroscopic and photometric datasets. Specifically, we included the spectroscopic data from the \emph{ShaSS} survey \citep{Haines2018}, the GLADE+ catalog \citep[][VizieR catalog VII/291]{Dalya2022}, and the DESI catalog \citep{Wen2024}.

Although redshift counterparts were identified for eight sources (Table~\ref{tbl:redshifts}, Appendix~\ref{SECT:AppendixB}), the available measurements suggest that the sample is not physically confined to a single structure.
While some sources are confirmed members of the main SSC core at $z \approx 0.048$ \citep{Bardelli1998,Proust2006a}, others appear associated with higher-redshift overdensities, including structures at $z \approx 0.10 - 0.15$ and the more distant sheets at $z \approx 0.17 - 0.20$ \citep{Quintana2000,Proust2006a,Merluzzi2015, Haines2018, Higuchi2020}.
The fact that these sources, despite being stratified across a wide range of redshifts ($0.048 \le z \le 0.20$), exhibit consistent projected $PA$s aligned with the X-ray filament is a noteworthy observation that we will discuss in Section~\ref{sec:discussion}.

\section{Optical counterparts and host galaxy alignment}
\label{sec:optical}

The optical counterparts, primarily identified in VST $r_{SDSS}$-band images \citep{Merluzzi2015}, suggest a possible trend in the relative orientation of radio jets and host-galaxy morphology. In particular, when an optical counterpart is resolved and appears elongated, its major axis tends to be approximately orthogonal to the radio jet.

This orthogonality is observed in four aligned sources: ID 34840 ($z_{spec}=0.043$, Figure~\ref{img:not_stacked_north_sample_B}), ID 35265 ($z_{spec}=0.0483$, Figure~\ref{img:not_stacked_north_sample_all}; despite its complex radio morphology, the stellar component retains an axis orthogonal to the primary jet), ID 35311 ($z_{phot}=0.24$, Figure~\ref{img:stacked_north_sample_all}), and ID 53568 ($z_{spec}=0.0497$, Figure~\ref{img:stacked_south_sample_all}). For ID 19874 (Figure~\ref{img:stacked_north_sample_all}), the jets are oriented at $\sim45^{\circ}$ to both the semimajor and semiminor axes. Only ID 53683 (Figure~\ref{img:NOT_stacked_south_sample_A}) shows a semimajor axis apparently aligned with the jets and parallel to the western filament. For the remaining sources, the optical orientation is unclear or the signal-to-noise ratio is too low for a reliable estimate.

Although the sample is too small for definitive conclusions, our findings suggest that radio jets preferentially align with host-galaxy minor axes. This is consistent with SMBH spins aligned with the angular momentum of the host galaxies. We discuss these implications in Section~\ref{sec:discussion}.

\section{Statistical significance of jet alignment}
\label{sec:stats}

To assess the statistical significance of the observed alignments, we employ analytical, numerical, and circular-statistical methods. Since each test relies on its own assumptions and is subject to distinct methodological biases, no single approach is fully comprehensive. Taken together, these complementary methods provide a more reliable range of plausible significance levels for the observed alignments.
  
We first calculate the binomial probability of observing a given number of aligned sources within an angular window $\Delta\theta$, both assuming a fixed reference orientation set by the DM filament $PA$ ($\mathrm{fixed}\ \theta$, $\theta=115^{\circ}$) and for all possible orientations ($\mathrm{Any}\ \theta$).
We then perform $10^6$ Monte Carlo (MC) realizations, perturbing each jet $PA$ according to its uncertainty. Accounting for measurement noise naturally reduces the apparent significance, but makes the assessment statistically more robust.
Finally, we apply the Rayleigh and $V$ tests for directional uniformity as an independent check that does not require a predefined angular window.

\subsection{Binomial approach}
\label{sec:stats_Binomial}
Assuming a random distribution of the $PA$s of the jets, the probability $p$ of a jet falling within a specific angular window $\Delta\theta$ from a predefined direction is $p = \Delta\theta / 180^{\circ}$.
The probability of finding exactly $k$ out of $n$ jets aligned along the same direction is given by:
\begin{equation}
P(k; n, p, \text{fixed}\ \theta) = \binom{n}{k} p^k (1-p)^{n-k}.
\end{equation}
Therefore, the cumulative probability $P$ of finding at least $k$ aligned sources out of $n$ is given by:
\begin{equation}
\label{EQ:P_fix}
P(\ge k; n, p, \text{fixed}\ \theta) = \sum_{i=k}^{n} \binom{n}{i} p^i (1-p)^{n-i}.
\end{equation}

This calculation assumes that the preferred alignment direction is specified a priori (i.e., the $PA$ of the Shapley filament). On the other hand, for a more conservative assessment of the significance of our observations, it is essential to evaluate the case in which the direction were not predefined (i.e., if the similar orientation of the filament is only incidental). In this case, the probability of finding at least $k$ sources whose $PA$s lie within a common angular window cannot be described by a simple binomial tail because the central direction of the window is allowed to vary over the full $180^{\circ}$ range \citep[look-elsewhere effect,][]{Gross2010}.

Following the scan statistics approach \citep{Naus1982,Glaz2013}, we can adopt the $PA$ of any one of the jets as the reference orientation. In this case, the probability that exactly $k-1$ of the remaining $n-1$ sources fall within the same angular window is given by the binomial distribution:

\begin{equation}
P_{\rm local} \simeq \binom{n-1}{k-1} p^{k-1} (1-p)^{n-k}.
\end{equation}

Given the number of effective directions $N_{\rm eff}\simeq n $, under the condition that such alignments are rare (i.e., $N_{\rm eff} P_{\rm local}\ll 1$), the probability that an alignment occurs for any direction can be approximated as $P_{\rm global} \simeq N_{\rm eff} \, P_{\rm local}$. This yields:

\begin{equation}
\label{EQ:P_any}
P(\ge k;n,p,\mathrm{any}\ \theta) \approx n \binom{n-1}{k-1} p^{k-1} (1-p)^{n-k}.
\end{equation}

We first evaluated the significance of the alignment observed in the northern and southern ``core'' samples, selected using restrictive morphological and size criteria. To assess whether the measured significance could be driven by the selection itself, we repeated the analysis using the more inclusive ``extended'' samples and their combination. This comparison tests whether including additional sources weakens the alignment signal (i.e., increases the p-value): stable or higher significance in the more inclusive samples would argue against the result being an artifact of sample selection. The scenarios and their corresponding probabilities are summarized in Table~\ref{tab:stats_final}.\footnote{The significances reported in this table are computed using the Python scripts available at \url{https://github.com/Ibaronch/Papers} (compute\_probabilities2.py).} For each scenario, we report the weighted mean of the jet $PA$ measurements listed in Table~\ref{tbl:redshifts}.

\begin{itemize}

\item {Scenario A (core north sample): Considering the six sources in the northern ``core'' sample aligned within a window of $20^{\circ}$ ($p=0.111$), with $n=6$ and $k=6$, we find $P(\text{fixed}\ \theta) \approx 1.88 \times 10^{-6}$ ($\sim 4.6\sigma$), whereas for an arbitrary direction, $P(\text{any}\ \theta) \approx 1.02 \times 10^{-4}$ ($\sim 3.7\sigma$). The weighted average $PA$ of the six selected sources is $\langle PA_\text{core north}\rangle=109.5^{\circ} \pm 2.8^{\circ}$.}

\item {Scenario B (extended north sample): Considering a total of 13 selected sources in the north (excluding ID 35348, with v-shaped jets), with 11 sources aligned within $40^{\circ}$ ($p = 0.222$), we obtain $P(\text{fixed}\ \theta) \approx 3.23 \times 10^{-6}$ ($\sim 4.5\sigma$) and $P(\text{any}\ \theta) \approx 1.52 \times 10^{-4}$ ($\sim 3.6\sigma$). In this case, the weighted average $PA$ of the 11 selected sources is $\langle PA_\text{extended north}\rangle=117.3^{\circ} \pm 1.9^{\circ}$.}

\item {Scenario C (core south sample): Considering the five southern sources meeting selection criteria (including ID 53742), four are aligned within an observed spread of $\Delta\theta = 10^{\circ}$ ($p = 0.056$), yielding $P(\text{any}\ \theta) \approx 3.24 \times 10^{-3}$ ($\sim 2.7\sigma$). Under the ``fixed $\theta$'' hypothesis, the angular window is expanded to $\Delta\theta = 20^{\circ}$ ($p = 0.111$) to encompass the filament $PA$ ($115^{\circ}$), giving $P(\text{fixed}\ \theta) \approx 6.94 \times 10^{-4}$ ($\sim 3.2\sigma$). The weighted average $PA$ of the four stacked sources is $\langle PA_\text{core south}\rangle=127.3^{\circ} \pm 3.2^{\circ}$.}

\item {Scenario D (extended south sample): Considering the full southern population, ten out of 13 sources (excluding ID 34624) align within an observed spread of $\Delta\theta = 15^{\circ}$ ($p = 0.083$), $P(\text{any}\ \theta) \approx 4.27 \times 10^{-7}$ ($\sim 4.9\sigma$). To encompass the filament $PA$ under the ``fixed $\theta$'' hypothesis, $\Delta\theta$ is expanded to $20^{\circ}$ ($p = 0.111$), resulting in $P(\text{fixed}\ \theta) \approx 5.96 \times 10^{-8}$ ($\sim 5.3\sigma$). The weighted average $PA$ of the ten selected sources is $\langle PA_\text{extended south}\rangle=128.5^{\circ} \pm 2.3^{\circ}$.}

\item {Scenario E (global significance): Considering the northern and southern alignments as independent local phenomena, the global (``joint'') probability of observing both of them is the product of the probabilities obtained in scenarios B and D: $P_{\text{fixed\_global}} \approx 1.93 \times 10^{-13}$ ($\sim 7.3\sigma$). Accounting for the look-elsewhere effect in both regions, the joint probability is $P_{\text{any\_global}} \approx 6.51 \times 10^{-11}$ ($\sim 6.4\sigma$). If we more conservatively do not consider the alignments in the northern and southern sectors as independent phenomena, the ``combined'' sample of 26 sources with 21 aligned within a $40^{\circ}$ window ($p = 0.222$), leads to $P_{\text{fixed}} \approx 3.83 \times 10^{-10}$ ($\sim 6.2\sigma$) and $P_{\text{any}} \approx 3.39 \times 10^{-8}$ ($\sim 5.4\sigma$). The weighted average $PA$ of the 21 selected sources is $\langle PA_\text{global}\rangle=121.7^{\circ} \pm 1.5^{\circ}$.}

\end{itemize}

We verified the accuracy of Equations~(\ref{EQ:P_fix}) and~(\ref{EQ:P_any}) across the different scenarios by comparing the analytically computed probabilities with the frequencies derived from a first series of Monte Carlo (MC) simulations under the null hypothesis of randomly oriented jets (Table~\ref{tab:stats_final_MC} in Appendix~\ref{SECT:AppendixD}). In each of the $5\times10^{9}$ realizations, $n$ values of $PA$ were drawn from a uniform distribution in the interval $[0^{\circ},180^{\circ}]$. First, for a fixed reference direction, we counted the number of jets falling within the angular interval $\Delta\theta$, yielding the corresponding binomial probability. Second, to account for the look-elsewhere effect, we searched for any angular interval $\Delta\theta$ containing at least $k$ sources, irrespective of its orientation. This was implemented by sorting the $PA$s and testing all possible groups of $k$ consecutive sources.

\begin{table*}[h]
\small
\caption{\label{tab:stats_final}Theoretical alignment significance, for different scenarios, from binomial computation.}
\centering
\begin{tabular}{|lcccc|cc|cc|}
\hline \hline
Scenario & $n$ & $k$ &  $\Delta\theta^a$    & $\langle PA \rangle$ &  Theoretical            & Significance     & Theoretical           & Significance \\
         &     &     & [deg]                &      [deg]$^b$       & $P$(fixed $\theta$)$^c$ & (fixed $\theta$) & $P$(any $\theta$)$^d$ &  (any $\theta$) \\
\hline
Core north (A)      & 6  & 6  & 20          & $109.5^{\circ} \pm 2.8^{\circ}$ & $1.88 \times 10^{-6}$ & $4.62\sigma$ & $1.02 \times 10^{-4}$ & $3.71\sigma$ \\
Extended north (B)  & 13 & 11 & 40          & $117.3^{\circ} \pm 1.9^{\circ}$ & $3.23 \times 10^{-6}$ & $4.51\sigma$ & $1.52 \times 10^{-4}$ & $3.61\sigma$ \\
Core south (C)      & 5  & 4  & 10 (20$^a$) & $127.3^{\circ} \pm 3.2^{\circ}$ & $6.94 \times 10^{-4}$ & $3.20\sigma$ & $3.24 \times 10^{-3}$ & $2.72\sigma$ \\
Extended south (D)  & 13 & 10 & 15 (20$^a$) & $128.5^{\circ} \pm 2.3^{\circ}$ & $5.96 \times 10^{-8}$ & $5.29\sigma$ & $4.27 \times 10^{-7}$ & $4.92\sigma$ \\
\hline
\textbf{Global (joint)} $^e$ & - & - & - & -                            & $1.93 \times 10^{-13}$ & $\mathbf{7.26\sigma}$ & $6.51 \times 10^{-11}$ &  $\mathbf{6.43\sigma}$ \\
\textbf{Global (combined)} $^f$ & 26 & 21 & 40 & $121.7^{\circ} \pm 1.5^{\circ}$ & $3.83 \times 10^{-10}$ &  $\mathbf{6.15\sigma}$ & $3.39 \times 10^{-8}$ &  $\mathbf{5.40\sigma}$ \\
\textbf{Global (combined) MC} $^g$ & 26 & 21 & - & - & $4.46_{0.23}^{91.9} \times 10^{-8}$ &  $\mathbf{5.35_{4.77}^{5.86}\sigma}$ & $2.91_{0.18}^{50.6} \times 10^{-6}$ &  $\mathbf{4.53_{3.89}^{5.09}\sigma}$ \\
\hline

\end{tabular}
\tablefoot{ \\
  \tablefoottext{a}{$\Delta\theta$ represents the observed angular spread of the jet $PA$s, adopted under the look-elsewhere assumption (``any $\theta$''). Under the ``fixed $\theta$'' hypothesis, the angular window must include the filament $PA$, requiring it to be widened to encompass it. This expanded $\Delta\theta$ is reported in parentheses when needed.}
\tablefoottext{b}{ For each scenario, the average $\langle\mathrm{PA}\rangle$ was computed as a weighted mean with weights $w_i = 1/\sigma_i^2$, where the individual uncertainties reported in Table~\ref{tbl:redshifts} are treated as Type B rectangular errors and converted to Gaussian standard uncertainties ($\sigma_i = a_i / \sqrt{3}$). The associated uncertainty on the average is given by $\sqrt{1 / \sum w_i}$. The Python script used for these calculations can be found at \url{https://github.com/Ibaronch/Papers} (compute\_PA2.py).}
\tablefoottext{c}{ Calculated using Equation~\ref{EQ:P_fix}.}
\tablefoottext{d}{ Calculated using Equation~\ref{EQ:P_any}.}
\tablefoottext{e}{ Calculated as the product of independent probabilities from Scenarios B and D, described in Section~\ref{sec:stats}. }
\tablefoottext{f}{ Calculated by pooling all sources in scenarios B and D, described in Section~\ref{sec:stats}.}
\tablefoottext{g}{ Calculated from $10^6$ MC realizations in which the individual jet $PA$s were perturbed according to their measurement uncertainties (treated as Type B errors). In each realization, the angular window width $\Delta\theta$ was dynamically tailored to the perturbed $PA$s distribution. Reported values represent the median, with the 16th and 84th percentiles indicated as lower and upper bounds.}
}

\end{table*}

The binomial significances remain dependent on the angular window width ($\Delta\theta$) and center ($\theta$), even after accounting for the look-elsewhere effect associated with the choice of central orientation. The windows were chosen to reflect the observed dispersion of the bulk of the $PA$s in each subsample, with the few outliers clearly separated from the main distribution, lying nearly orthogonal to it. Nevertheless, this physically motivated choice implicitly favors higher estimated significances: a narrower window would exclude some aligned jets, whereas a wider one would retain the same number of sources but increase the background probability, thereby reducing the overall significance.

\begin{figure}[!ht]
\centering
\resizebox{\hsize}{!}{\includegraphics{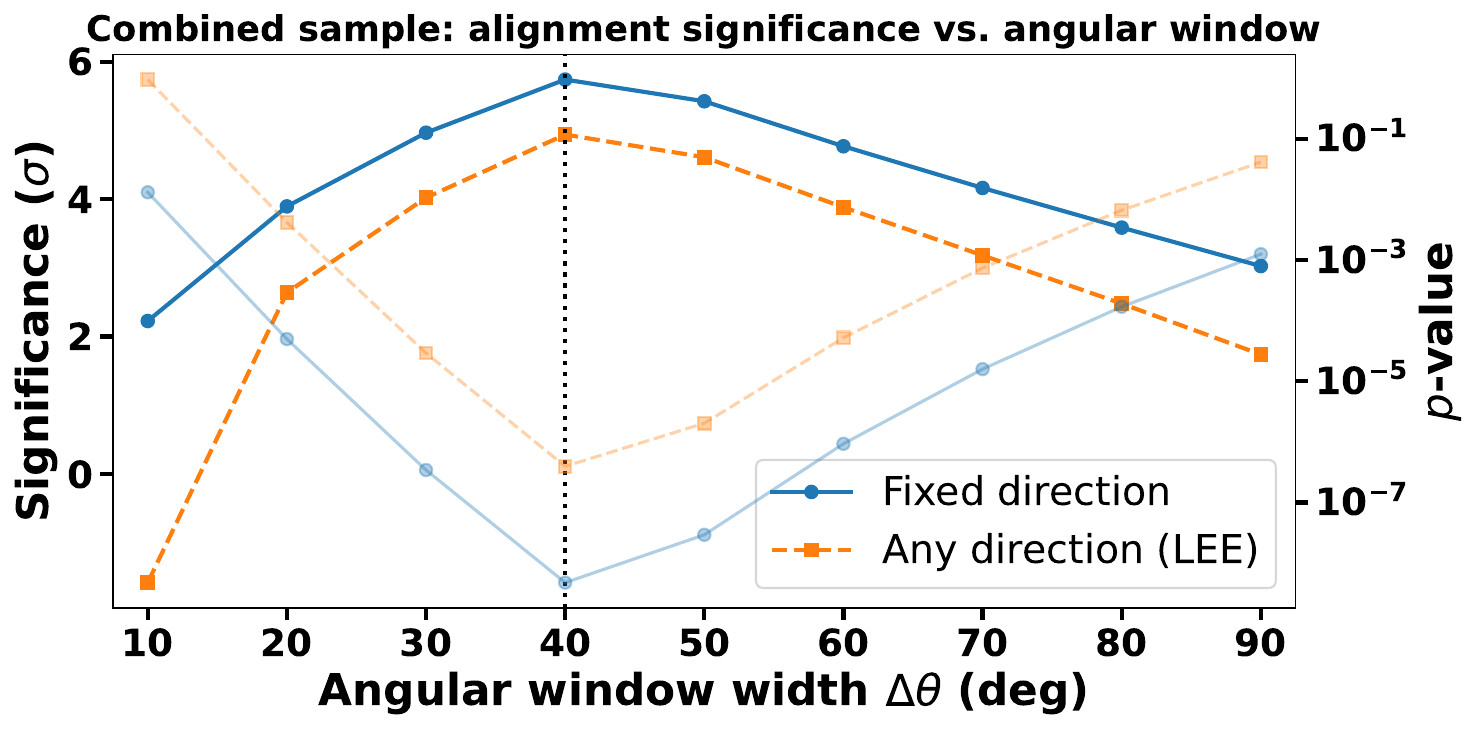}}
\caption{Binomial alignment significance as a function of the angular window considered. The right axis shows $P(\mathrm{fixed}\ \theta)$ and $P(\mathrm{any}\ \theta)$ versus the angular-window width $\Delta\theta$, as faint blue and orange lines, respectively. The corresponding significances (left axis) are shown as saturated lines. Blue lines represent binomial estimates for a fixed central orientation aligned with the filament ($\theta_0 = 115.0^{\circ}$), while dashed orange lines account for the look-elsewhere effect. The vertical dotted line marks the significance peak at $\Delta\theta = 40^{\circ}$, corresponding to the characteristic dispersion of the jets. Beyond this width, no additional sources enter the window until $\Delta\theta\sim100^{\circ}$, and the significance gradually reduces. }
\label{img:significance_vs_delta_theta}
\end{figure}

Figure~\ref{img:significance_vs_delta_theta} shows the computed significance and corresponding p-values as functions of the angular window width $\Delta\theta$ for the global ``combined'' scenario. For $\Delta\theta \lesssim 40^{\circ}$, widening the window captures aligned sources faster than expected for an isotropic distribution, increasing the significance. Beyond $\sim40^{\circ}$, no additional sources enter the window until $\Delta\theta\sim100^{\circ}$, while the background probability increases, causing the significance to decline. The peak near $40^{\circ}$ reflects the intrinsic dispersion of the jet alignment around the cosmic filament (which defines the window center in this test). The maximum significance in this test is lower than in our binomial calculation because here the window is centered on the filament $PA$ ($115^{\circ}$), rather than on the center of the observed jet-$PA$ range ($120^{\circ}$).

\subsection{MC simulations based on a binomial approach}
\label{sec:stats_Binomial}

\begin{figure}[!ht]
\centering
\resizebox{\hsize}{!}{\includegraphics{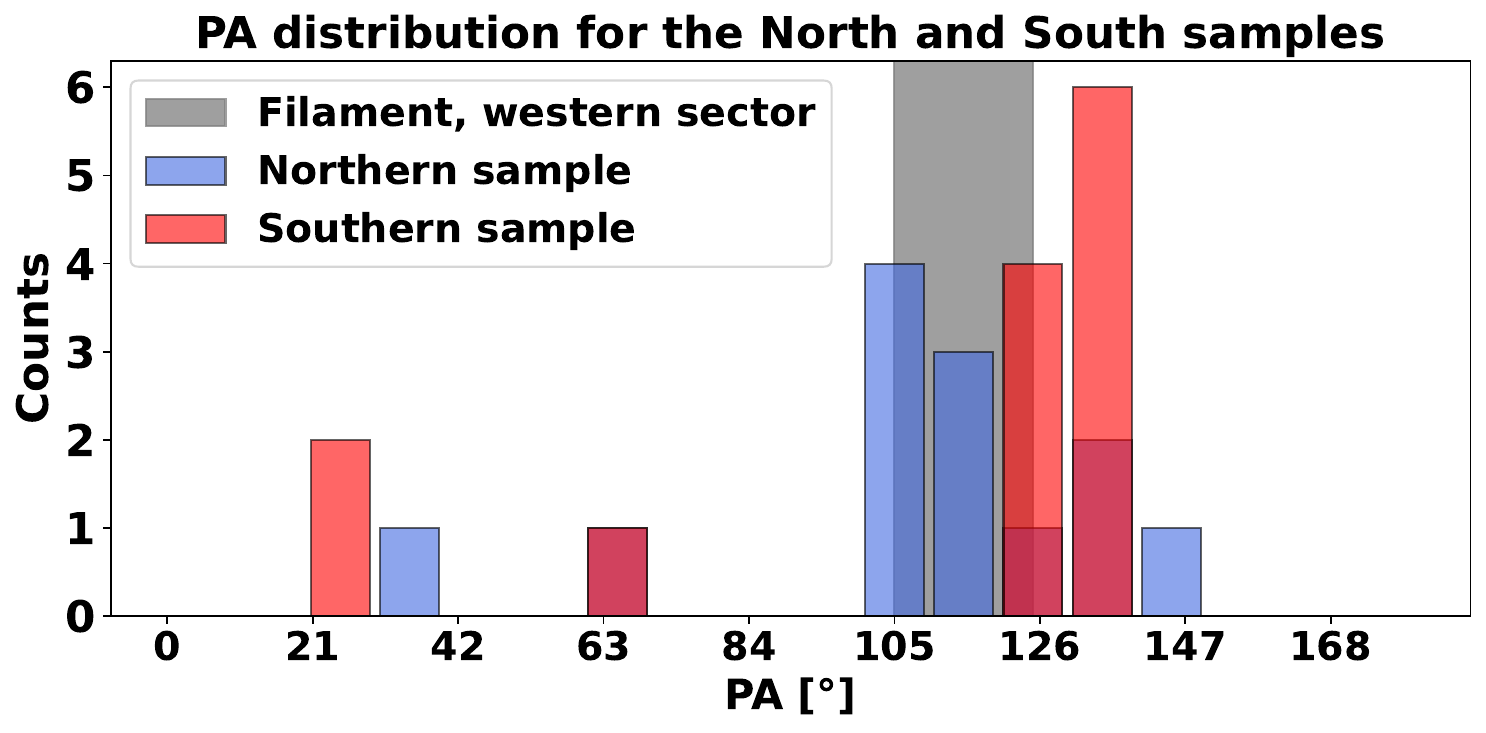}}
\caption{Histogram of the $PA$s with a bin width of $10^{\circ}$ comparing the northern sample (blue bars, including both stacked and non-stacked sources) and the southern sample (red bars). Sources with undefined or ambiguous $PA$s are omitted. The gray shaded region marks the $115^{\circ} \pm 10^{\circ}$ range, representing the alignment of the western sector of the filament (see Figure~\ref{img:filament_alignement}).}
\label{img:Position_angles_hist}
\end{figure}

\begin{figure}[!ht]
\centering
\resizebox{\hsize}{!}{\includegraphics{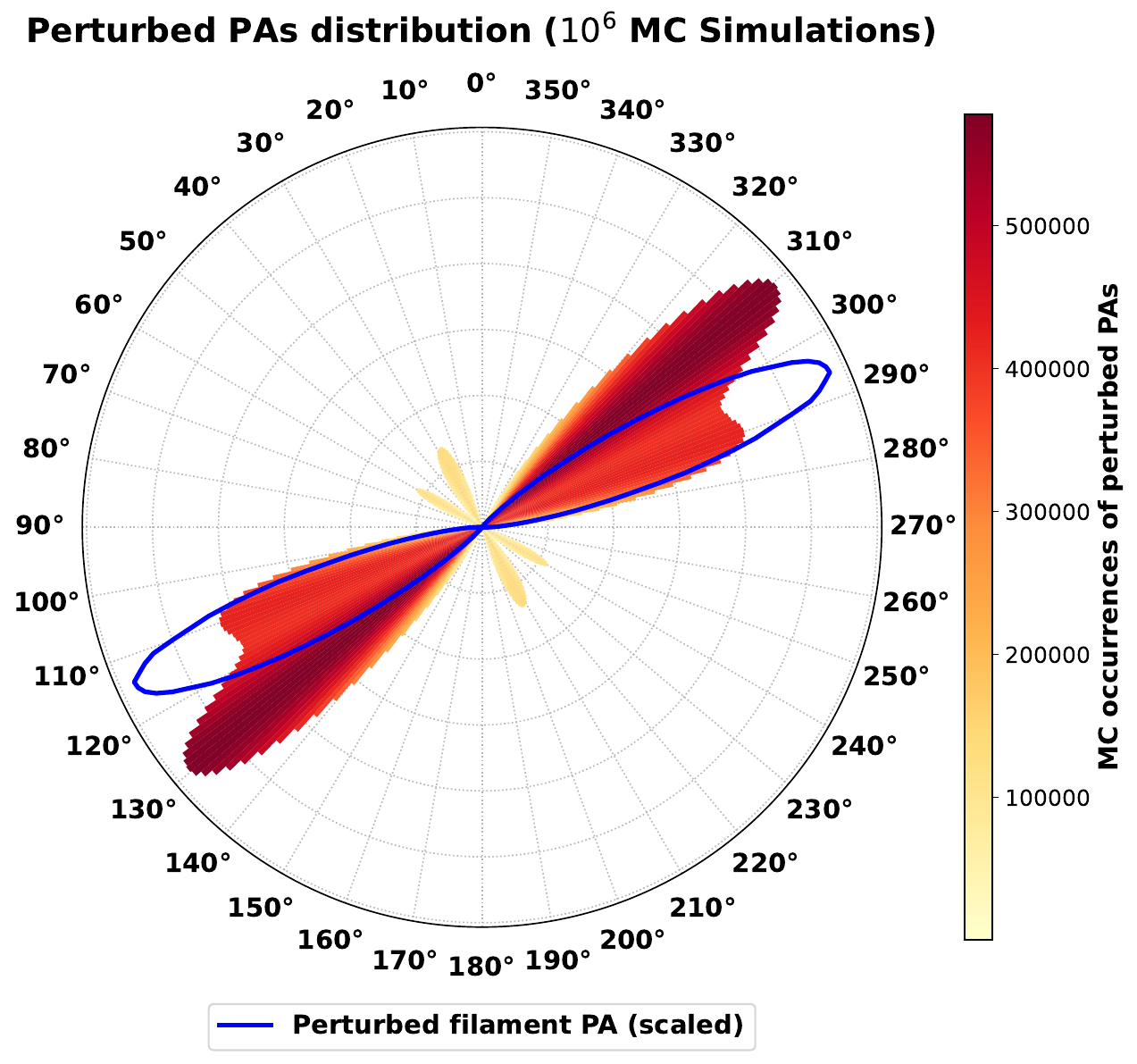}}
\caption{Polar distribution of perturbed $PA$s from $10^6$ MC simulations of the combined jet sample. The color-coded histogram shows occurrences per angular bin, ranging from light yellow (low frequency) to dark red (high frequency). The solid blue line represents the perturbed filament orientation ($\theta_0$), with its area scaled to match the $PA$ histogram's peak height for comparison. }
\label{img:combined_polar_distribution_MC}
\end{figure}

Figure~\ref{img:Position_angles_hist} displays the $PA$ histograms for the northern and southern ``extended'' samples. The two distributions exhibit a slight difference in their mean position angles ($117.3^{\circ} \pm 1.9^{\circ}$ versus $128.5^{\circ} \pm 2.3^{\circ}$), but both are roughly aligned with the $PA$ of the filament ($115^{\circ} \pm 10^{\circ}$).
The polar diagram in Figure~\ref{img:combined_polar_distribution_MC} shows the same $PA$ distribution perturbed according to the estimated uncertainties, based on a second series of $10^6$ MC realizations. The similarly perturbed filament $PA$ is also shown.

For this computation, the $PA$ uncertainties reported in Table~\ref{tbl:redshifts} were treated as ``Type B'' uncertainties \citep{JCGM1002008}. Unlike ``Type A'' uncertainties derived statistically from repeated measurements, ``Type B'' uncertainties apply when a quantity is known only within a specific interval. In our sample, an exact $PA$ cannot be uniquely defined without introducing methodological biases: for instance, the central source is not always identifiable, and using the brightest pixel of the outflow rather than its geometric center yields systematically different orientations. We therefore assume that the true jet axis lies (with uniform probability distribution) between the minimum and maximum angles $\mathrm{PA}\mathrm{min}=PA-a$ and $\mathrm{PA}\mathrm{max}=PA+a$. To account for these uncertainties in our MC simulations, we converted them into Gaussian standard deviations. Assuming a rectangular probability distribution, the standard uncertainty is $\sigma_\mathrm{PA} \simeq a/\sqrt{3}$, where $a$ is the interval half-width \citep{JCGM1002008}.

The cumulative $PA$ distribution from the MC simulations exhibits two prominent peaks at distinct average angles, with the filament lying approximately between them. The five misaligned jets produce secondary, less significant distributions, nearly orthogonal to the primary peaks.

In each MC realization, the angular-window width $\Delta\theta$ has been dynamically adjusted to the range of perturbed $PA$s\footnote{Given their nearly orthogonal orientations, the five unaligned sources almost never fall within the window defined by the perturbed $PA$s of the remaining aligned jets.}. This accounts for variability across realizations and mitigates the potential selection bias associated with fixing $\Delta\theta$ a priori at a value that maximizes the analytical binomial significance.

For the Global (Combined) scenario, in the series of MC simulations just described, we obtain\footnote{Under the fixed-$\theta$ assumption, the adopted angular window that always encloses the filament $PA$ ($\theta_{0}$), which has been perturbed in the same manner as the jet $PA$s.}: $P(\mathrm{fixed}\ \theta)=4.46^{91.9}_{0.23} \times 10^{-8}$ ($5.35^{5.86}_{4.77} \sigma$) and $P(\mathrm{Any}\ \theta)=2.91^{50.6}_{0.18}\times 10^{-6}$ ($4.53^{5.09}_{3.89}\sigma$), where reported values represent the median with the $16^\mathrm{th}$ and $84^\mathrm{th}$ percentiles as bounds\footnote{The significances obtained in the unperturbed and static $\Delta\theta$ binomial analysis fall between the fifth and sixth highest percentiles of the MC simulation results.}. 
As expected, these median values are slightly lower than the static analytical values in Table~\ref{tab:stats_final}. This occurs because introducing perturbed realizations naturally smooths out the signal while accounting for the observational uncertainty. Therefore, we adopt this result as a more realistic and conservative statistical estimate of the significance.

\subsection{Circular statistics: Rayleigh and V-tests}

As a complementary, window-independent assessment, we apply continuous circular-statistical tests \citep[for a comparison of common methods, see][]{Landler2018}. These tests probe angular concentration differently from the binomial approach, providing an independent consistency check rather than an alternative measure of alignment significance.

For $n$ sources, the standard Rayleigh test evaluates the null hypothesis of a uniform circular distribution against unimodal clustering along an unspecified direction. Defining the vector components as $C = \sum_{i=1}^{n} \cos\phi_i$ and $S = \sum_{i=1}^{n} \sin\phi_i$, the mean resultant length $\bar{R}$ is given by $\bar{R} = (1/n)\sqrt{C^2 + S^2}$.

The Rayleigh statistic $Z=n\bar{R}^2$ quantifies directional concentration. Following \cite{Mardia1975} and \cite{Wilkie1983}, under $H_0$, the upper-tail probability of obtaining a resultant length at least as large as observed is approximated by:
\begin{equation}
P_{\text{Rayleigh}} \approx \exp(-Z) \left[ 1 + \frac{2Z - Z^2}{4n} - \frac{24Z - 132Z^2 + 76Z^3 - 9Z^4}{288 n^2} \right].
\end{equation}

The Rayleigh test assesses unimodal departures from circular uniformity without requiring a preferred direction. However, the western segment of the Shapley X-ray filament defines a physically motivated reference $PA$ ($\theta_0 = 115.0^{\circ}$). We therefore also apply the modified Rayleigh ($V$-) test \citep{Durand&Greenwood1958,Zar1999}, which tests for directional concentration around $\theta_0$. Defining the relative angular offsets as $\alpha_i = 2(\theta_i - \theta_0)$, the test statistic, its standardized form, and the corresponding one-tailed $p$-value are given by:
\begin{equation}
V = \sum_{i=1}^{n} \cos\alpha_i, \qquad
u = V\sqrt{\frac{2}{n}}, \qquad
P_V = 1-\Phi(u),
\end{equation}
where $\Phi$ is the standard normal cumulative distribution function and $u$ asymptotically follows a standard normal distribution under $H_0$, directly providing the Gaussian-equivalent significance in units of $\sigma$.

The results of both circular tests across all scenarios are reported in Table~\ref{tab:circular_stats}. Both confirm significant jet alignment ($3.83\sigma$ for the Rayleigh test and $4.16\sigma$ for the $V$-test in the ``Global (combined)'' scenario), without relying on ad hoc angular windows. However, the significances are lower than those obtained with the binomial approach.

\begin{table*}[t]
\small
\caption{\label{tab:circular_stats} Theoretical alignment significance for different scenarios from circular statistics (Rayleigh test and $V$-test with predefined reference direction $\theta_0 = 115.0^{\circ}$).}
\centering
\begin{tabular}{|l c | c c | c c|}
\hline\hline
Scenario           & $n$ & $V$-test $p$-value  & Significance  & Rayleigh                       & Significance   \\
                   &     & $P_V$               & ($\sigma$)    & $P_{\text{Rayleigh}}$             &   ($\sigma$)   \\  
\hline                                                                                                                 
Core north (A)     & 6  & $4.84 \times 10^{-4}$ & $3.30\sigma$  & $3.01 \times 10^{-4}$          & $3.43\sigma$  \\  
Extended north (B) & 13 & $3.55 \times 10^{-4}$ & $3.39\sigma$  & $1.68 \times 10^{-3}$          & $2.93\sigma$  \\  
Core south (C)     & 4  & $6.51 \times 10^{-3}$ & $2.48\sigma$  & $8.33 \times 10^{-3}$          & $2.39\sigma$  \\  
Extended south (D) & 13 & $6.32 \times 10^{-3}$ & $2.49\sigma$  & $1.28 \times 10^{-2}$          & $2.23\sigma$  \\  
\textbf{Global (combined)}               & 26 & $1.61 \times 10^{-5}$ & $\mathbf{4.16\sigma}$ & $6.46 \times 10^{-5}$ & $\mathbf{3.83\sigma}$ \\
\textbf{Global (combined) MC-est. a $^a$} & 26  & $2.23_{0.47}^{8.86}\times 10^{-6}$ & $\mathbf{4.59_{4.29}^{4.91} \sigma}$ & $2.86_{0.47}^{16.1}\times 10^{-6}$ & $\mathbf{4.54_{4.16}^{4.91} \sigma}$  \\
\textbf{Global (combined) MC-est. b $^b$} & 26  & $8.02_{0.99}^{92.3}\times 10^{-6}$ & $\mathbf{4.31_{3.74}^{4.76} \sigma}$ & $7.34_{0.90}^{48.3}\times 10^{-6}$ & $\mathbf{4.33_{3.90}^{4.77} \sigma}$  \\
\hline
\end{tabular}
\tablefoot{ \\
  \tablefoottext{a}{ Continuous circular tests underestimate the true statistical significance when secondary preferential directions exist, as the distribution of the $PA$s suggests (see Figure~\ref{img:Position_angles_hist} and~\ref{img:combined_polar_distribution_MC}). The estimate ``a'' has been obtained by replacing the $PA$s of the five not-aligned sources with random - but still not aligned - values (i.e., $\theta_i$ outside the range $100^{\circ}-140^{\circ}$) and repeating the test $10^5$ times.  }
  \tablefoottext{b}{ The estimate ``b'' has been obtained by performing $10^6$ MC simulations in which the original $PA$s of the jets are perturbed using the associated uncertainties \citep[following][these are the type ``B'' errors reported in Table~\ref{tbl:redshifts} divided by $\sqrt{3}$]{JCGM1002008}. The correction described in the note ``a'' is also incorporated.  }
  }
\end{table*}

This lower significance is likely due to the fact that circular statistics tend to underestimate the strength of a primary alignment in the presence of secondary alignment directions. Since $PAs$ are axial (defined modulo $180^{\circ}$), all angles are doubled ($\phi_i=2\theta_i$) before computing vector sums, mapping them onto a full $360^{\circ}$ circle. Consequently, orientations separated by $\sim90^{\circ}$ in real space map to antipodal vectors ($180^{\circ}$ apart) on the circle, causing partial vector cancellation ($R\to0$). The resulting test statistics are therefore conservative and tend to reduce the significance of the main alignment.

As shown by the $PA$ distribution (Figures~\ref{img:Position_angles_hist} and~\ref{img:combined_polar_distribution_MC}), five sources form a secondary cluster roughly orthogonal to the filament axis, with $PA$s between $25^{\circ}$ and $60^{\circ}$. Although too few to draw conclusions about their own alignment, these sources have a non-negligible impact on circular statistics. To quantify this vector cancellation, we performed $10^5$ MC realizations, replacing their $PA$s with values drawn uniformly outside the primary alignment window ($100^{\circ} < \theta < 140^{\circ}$, with a $10^{\circ}$ buffer).
In this idealized scenario, the significances increase ($P_{\text{Rayleigh}} = 2.86_{0.47}^{16.1}\times 10^{-6}$, $4.54_{4.16}^{4.91}\sigma$ and $P_V = 2.23_{0.47}^{8.86}\times 10^{-6}$, $4.59_{4.29}^{4.91}\sigma$), confirming that orthogonal sources  systematically depress circular-test significance.

As for the binomial analysis, we performed $10^6$ MC simulations, perturbing the jet $PA$s according to their associated uncertainties, obtained by dividing the ``Type B'' errors reported in Table~\ref{tbl:redshifts} by $\sqrt{3}$ \citep{JCGM1002008}. We also incorporated the correction due to orthogonal sources, described above. This yields $P_{\text{Rayleigh}} = 7.34_{0.90}^{48.3}\times 10^{-6}$ ($4.33_{3.90}^{4.77}\sigma$) and $P_V = 8.02_{0.99}^{92.3}\times 10^{-6}$ ($4.31_{3.74}^{4.76}\sigma$).

As our MC simulations show, the Rayleigh and $V$-tests are designed to detect unimodal departures from uniformity and are therefore less suited to the apparently multi-component alignment visible in Figures~\ref{img:Position_angles_hist} and~\ref{img:combined_polar_distribution_MC}. Our primary hypothesis concerns alignment with the filament axis within a finite angular tolerance, which is more directly tested by the binomial approach, particularly under the fixed-$\theta$ configuration. We therefore use circular tests as complementary, window-independent tests of non-uniformity rather than as alternatives to the binomial approach.

For the Global (combined) scenario, the alignment significance ranges from $3.83\sigma$, obtained with the Rayleigh test, to $6.15\sigma$, derived from the analytical binomial calculation under the ``fixed-$\theta$'' assumption. These are not formal bounds but illustrate the range obtained with different statistical approaches. We adopt as our fiducial estimate the median significance from the binomial MC simulations under the ``Any $\theta$'' configuration ($4.53\sigma$), which accounts for observational uncertainties and the look-elsewhere effect. This estimate is also consistent with the circular-statistical results once the effect of the nearly orthogonal sources is accounted for.

\section{Discussion}
\label{sec:discussion}

The alignment of the radio AGN jet axes observed across the northern and southern sectors of the SSC poses a compelling interpretive challenge when reconciled with the available redshift data (Table~\ref{tbl:redshifts}).
On one hand, the projected spatial coherence is striking: all six selected sources in the northern sector exhibit a coherent orientation within an angular spread of $\sim 20^{\circ}$, while four out of five in the southern sector are aligned within $\sim 10^{\circ}$. As demonstrated in Table~\ref{tab:stats_final}, this alignment is remarkably robust; even when relaxing our ``core'' selection criteria, the statistical significance remains comparable in the northern sector and increases in the southern sector. Aggregating the two sky areas, 21 out of 26 sources are aligned within $\sim 35\text{--}40^{\circ}$, yielding a significance of $4.5^{6.1}_{3.8}$.

Conversely, this alignment must be interpreted in light of the ``stratified'' redshift distribution detailed in Table~\ref{tbl:clusters}. Our sample probes multiple redshift bins along the LoS: while sources such as ID 35265 ($z=0.0483$) are confirmed members of the local SSC core \citep{Bardelli1998}, others, such as ID~19874 and ID~53683, reside at significantly greater distances ($z \approx 0.16\text{--}0.20$). Such a redshift spread far exceeds the scale of peculiar velocities within the A3558 complex, suggesting that the coherence observed extends radially along the LoS.

If the higher-redshift sources ($z \approx 0.17 - 0.20$) indeed "mimic" the orientation of the local SSC filament ($z \approx 0.048$), it might suggest that the cosmic web in this sector of the sky maintains a degree of coherence over significant cosmological distances, across distinct virialized structures. In this perspective, our survey area effectively acts as a deep "pencil-beam" probe of the large-scale cosmic web. Nonetheless, due to the small number of available redshifts (only seven aligned objects with valid redshift measurements), drawing definitive conclusions about the 3D geometry is premature.

The region analyzed in our study hosts a significant "background echo" of the SSC. Previous wide-field surveys have identified major galaxy overdensities beyond $z \approx 0.048$ and up to $z\sim0.14$ and $0.25$ \citep{Quintana2000, Proust2006a, Proust2006b, Merluzzi2015, Haines2018, Higuchi2020}.

\cite{Ragone2006} characterized the large-scale density distribution and the mass-density fluctuations in this specific sector. Their analysis confirms that the Shapley region is embedded in a complex network of structures where the gravitational potential remains significant well beyond the local SSC core. 
In this framework, the radio sources identified at higher redshifts in our sample could be tentatively associated with these secondary layers of the cosmic web, which appear to be structurally aligned with the primary LSS of the SSC.
Importantly, rather than requiring the presence of a single, yet-undetected massive attractor beyond Shapley, the alignment can be the result of a statistically coherent anisotropic mass distribution aligned with the preferred direction defined by the local tidal field.

Recent dynamical reconstructions of the local velocity field further support this interpretation. In particular, \cite{Stiskalek2026} show that the Local Group peculiar velocity cannot be attributed to a single dominant attractor, including the SSC itself, but rather arises from a superposition of contributions across progressively larger scales. In their analysis, streamline convergence depends strongly on smoothing scale, with flow patterns transitioning from local structures (e.g., Virgo) through intermediate overdensities (e.g., Hydra--Centaurus) and extending toward Shapley only at large scales, without achieving a fully convergent basin within $\sim 155\,h^{-1}\,\mathrm{Mpc}$. This behavior is consistent with a nontrivial velocity field that remains sensitive to mass distributions beyond the canonical Shapley core.

A confirmed $PA$ coherence from the SSC up to redshift $z\sim0.2$ would correspond to about $\sim600$Mpc ($\sim1700$ Mpc if we consider the most distant source included in our analysis, at $z\sim0.5$). This size is well beyond the supposed homogeneity scale of the Universe \cite[$\sim$260-370Mpc, e.g.,][]{Ratra2008,Yadav2010}, but smaller than / comparable to other structures already discovered, such as the ``HBC Great Wall'' in the Hercules–Corona Borealis \cite[$\sim2000-3000$Mpc]{Horvath2015, Horvath2020}, the ``Giant RGB ring'' \cite[$\sim1720$Mpc,]{Balazs2015}, the ``U1.27 Huge-LQG'' \cite[$\sim1240$Mpc,][]{Clowes2013}, the ``Giant arc'' \cite[$\sim1000$Mpc]{Lopez2022}, the ``Correlated orientations of the Large quasar groups (LQG)'' \cite[$\sim$1600Mpc,][]{Friday2022}, the coherent quasars polarization \cite[$\sim1000$Mpc,][]{Hutsemekers1998,Hutsemekers2001,Hutsemekers2005} and many others \cite[for a reference, see Table 1 in][]{Lopez2022}.

Once again, it is crucial to clarify that this coherence does not imply the existence of a single, continuous, and virialized mega-structure, but that it can be interpreted as a "fossil" signature of the large-scale gravitational environment. In this scenario, the matter that now populates these distinct clusters or overdensities was originally part of a larger, near-homogeneous primordial superstructure. As cosmic evolution progressed, gravitational instability triggered the contraction of this matter along the preferred axes of the cosmic web. This directional collapse leads to a twofold result: the concentration of matter into discrete, virialized nodes (i.e., the clusters we observe today) and the simultaneous "emptying" of the surrounding space. 

This process of matter evacuation naturally accounts for the presence of significant underdense regions and discontinuities, such as the void identified behind the SSC in dynamical surveys \cite[in the range $z\sim0.05-0.07$, see Figures 6, 7, 8 in][]{Haines2018}. While the density field has evolved into a nonlinear and fragmented distribution of clusters and voids, the orientation of the supermassive black hole spin axes, traced here by the jet $PA$s, may have preserved a memory of the primordial angular momentum and tidal field that characterized the parent superstructure before its fragmentation. 

Since these regions lack the relaxed gravitational potential of massive clusters, the observed alignment, if confirmed, cannot be attributed to localized cluster-scale dynamics but it should be responding to the underlying DM web even before full virialization occurs. A possible physical mechanism for the morphological coherence observed across these distinct redshift shells lies in the framework of large-scale tidal alignment. According to models of structure formation based on tidal torque theory \citep[e.g.,][]{Codis2012, Dubois2014, GaneshaiahVeena2018}, the orientation of a galaxy's angular momentum, and potentially the spin axis of its central supermassive black hole, is influenced by the gravitational tidal field of the host LSS filament \citep{Tempel2013a}. Evidence of this theoretical transfer of angular momentum from filaments of the cosmic web to the individual galaxies has been observed over scales larger than those predicted by cosmological simulations \citep{Tudorache2025}.

Our preliminary analysis of the optical counterparts (Section \ref{sec:optical}) provides additional evidence in this direction. For most of the sources, the optical shape cannot be assessed or looks approximately circular. In one case (ID 53683) the semimajor axis of the optical counterpart is oriented in the same direction as the jets. In another case (ID 19874), the direction of the jets makes an angle of $\sim45^{\circ}$ with the semi-minor and major axes. However, for four sources (IDs 35311, 35265, 34840, and 53568), the radio jets appear aligned with the photometric minor axis of the host galaxy and are approximately orthogonal to the stellar major axis.

This  configuration supports a coherent accretion history where the angular momentum of the infalling gas, ultimately feeding the central engine, is coupled with the large-scale rotation of the stellar system.
In the context of the SSC core, this "ordered" configuration implies that the environmental influence of the cosmic filament extends down to subparsec scales. The consistency between the jet $PA$, the host galaxy orientation, and the large-scale X-ray structure suggests the possibility that the gravitational collapse and subsequent assembly of these systems occurred along preferential directions dictated by the cosmic web, preserving the dynamical memory of the filament across several orders of magnitude in scale.

The persistence of a parallel alignment between these jet axes and the filament backbone poses an interesting theoretical challenge. Numerical simulations \citep{Codis2012} predict a "spin-flip" transition from parallel to perpendicular orientation for DM halos exceeding a critical mass threshold of $M_{\rm crit} \sim 5 \times 10^{12} M_\odot$ at $z=0$, with a dependence on $z$ in the form $M_\mathrm{crit}(z)\propto (1+z)^{-2.5}$. Below this threshold, parallel alignment is preferred due to smooth gas accretion along the cosmic web; above it, major mergers occurring preferentially along filamentary axes flip the resulting spin to a perpendicular orientation \citep{West1994}. Since FR~II sources are typically hosted by massive elliptical galaxies, their parent halos would theoretically be expected to reside in the perpendicular-spin regime. 

Our findings could be tentatively reconciled with the tidal torque theory framework by considering the specific environmental context of these sources. As noted in Section  \ref{sec:cluster_env_and_z_distr}, the observed structures are largely non-virialized galaxy sheets rather than relaxed, massive clusters. Furthermore, the radio AGN in our sample are generally not the central engines of virialized clusters (BCGs), but rather peripheral members distributed along the putative filamentary spine. In such environments, the ratio between the visible stellar mass and the total halo mass might differ from that of central cluster galaxies. It is therefore possible that, despite their high stellar mass, these systems have not yet crossed the critical halo mass threshold required for the spin-flip to occur. In this scenario, directional gas accretion along cosmic streams would continue to dominate the angular momentum acquisition, preserving the primordial parallel alignment dictated by the LSS.

\section{Conclusions}
\label{sec:conclusions}

We detected a highly significant alignment ($4.5^{6.1}_{3.8}\sigma$) among 26 radio jets in the SSC region. A preliminary spectroscopic analysis (seven sources) suggests that background sources ($z\sim 0.17-0.20$) appear to share a common orientation with the main SSC core ($z\sim0.05$), reflecting the projection of multiple structures across cosmological scales, potentially exceeding the cosmic homogeneity scale.

The observed alignment is consistent with galaxy spin correlations induced by large-scale tidal fields, as predicted by tidal torque theory. However, while these results align with previous studies in the literature, conclusions regarding the 3D geometry remain tentative due to sparse redshift coverage and to the limited sample size.
Therefore, these results do not constitute a final proof of multi-scale alignment. They instead highlight an interesting consistency between the X-ray gas morphology, the galaxy dynamics, and the jet orientation that warrants further investigation with larger spectroscopic and high-resolution samples.

High-sensitivity observations over large fields from next-generation instruments, such as MeerKAT in South Africa \citep{Jonas2016}, are essential to providing the statistical robustness needed to map these orientations within the cosmic filamentary network. Such future analysis is also essential to determining if similar large-scale alignments represent a widespread feature of the cosmic web or just a peculiar characteristic of the Shapley region.

Nonetheless, these results underscore the importance of extended radio sources as effective tracers of the structure of the cosmic web. Even in the presence of a sparse spectroscopic coverage for all members, the statistical consistency of the jet orientations provides a powerful tool to identify large-scale structural coherence that transcends individual bound clusters.

Providing observational constraints on how the cosmic web shapes the orientation of galaxies has important implications for cosmology and galaxy formation; for instance, the study of these intrinsic alignments currently represents a key aspect of high-precision analyses in missions such as Euclid \citep{EuclidCollab2026_A,EuclidCollab2026_B,EuclidCollab2026_C,EuclidCollab2026_D}. While in this work we did not explore the details and the possible implications in this context, we emphasize that radio jet directions offer a complementary probe of these large-scale spatial correlations by tracing the spin axes of the active galaxies' SMBHs.

\section*{Data availability}
The Python scripts used for the computations described in this paper are available at \url{https://github.com/Ibaronch/Papers}, under the folder named "Baronchelli+2026".

\begin{acknowledgements}
  OMS's research is supported by the South African Research Chairs Initiative of the Department of Science and Technology and National Research Foundation (grant No. 81737).
\end{acknowledgements}

\bibliographystyle{aa}
\bibliography{biblio6}{}

\begin{appendix}
\nolinenumbers
\twocolumn
  
\section{Source extraction and catalogue construction}
\label{SECT:AppendixA}

Source extraction across the three partially overlapping MeerKAT maps was carried out utilizing a automated pipeline built around \texttt{SExtractor} \citep{Bertin1996}. A summary of the main \texttt{SExtractor} configuration (non-default) parameters is provided in Table~\ref{tab:sextractor_params}. The background is evaluated locally and the pipeline automatically selects the most appropriate Gaussian convolution kernel based on the characteristic FWHM of the images. The detection threshold is uniformly set to $1.5\sigma$ (\texttt{DETECT\_THRESH}), requiring a minimum number of connected pixels above this threshold (\texttt{DETECT\_MINAREA}) that is dynamically tied to the actual beam area.

\begin{table*}
\small
\caption{\label{tab:sextractor_params} Source extraction: non-default SExtractor parameters used$^{[a]}$. }
\centering
\begin{tabular}{lll}
\hline
\hline
SExtractor & Value & Units \\ 
parameter & & \\
\hline
filter & "Gauss" with size $\sim 2\times\mathrm{FWHM}$ & pixels \\
BACK\_SIZE       &$3\times\mathrm{FWHM}$ & pixels \\
BACK\_FILTERSIZE & 2 &  -  \\
DETECT\_MINAREA & $2\pi\left(\frac{\mathrm{FWHM_{a}}}{2.355}\right)\left(\frac{\mathrm{FWHM_{b}}}{2.355}\right)$ & pixels \\
DETECT\_THRESH   & 1.5 & $\sigma$ \\
ANALYSIS\_THRESH & 1.5 & $\sigma$ \\
PHOT\_AUTOPARAMS (KRON\_FACT)$^{[b]}$   & 2.5 & A\_IMAGE \\
PHOT\_AUTOPARAMS ($r_\mathrm{KRON min}$)$^{[c]}$   & $0.6\times\mathrm{FWHM}_\mathrm{PIX}$ & pixels \\
\hline
\end{tabular}
\tablefoot{ \\
\tablefoottext{a}{ After the initial SExtractor run, a successive refined source selection is performed as described in Appendix~\ref{SECT:AppendixA};}
\tablefoottext{b}{ KRON\_FACT is the size of the elliptical aperture used for the SExtractor "AUTO" photometry. It is expressed in units of A\_IMAGE and B\_IMAGE, i.e, the semimajor and semiminor axes of the elliptical representation of the source identified;}
\tablefoottext{c}{ $r_\mathrm{KRON min}$ represents the smallest size, in pixels, of the automatic aperture used.}
}

\end{table*}

The significance of each detections is computed as follows. Initially, a large sample of non-overlapping circular apertures are placed at random coordinates across the field. The diameter of these random apertures is fixed to match the beam size of each image ($D_{\text{beam}} = 2 \times \sqrt{\Omega_{\text{beam}} \cdot 4 \ln 2 / \pi}\sim13''$). The integrated flux is measured within each random aperture and the standard deviation of this flux distribution is computed ($\sigma_{\text{rand}}$). Similarly, for every source, the pipeline measures the integrated flux within a fixed circular aperture of the same beam-matched diameter ($F_{\text{aperture}}$). Finally, the significance is computed as $S/N=F_{\text{aperture}} / \sigma_{\text{rand}}$.

The final catalog is compiled by cross-matching the extractions from the three partially overlapping MeerKAT images. For each duplicate detection, the pipeline evaluates the local depth reached by accounting for both the local PB correction factor and the explicit image exposure time. The detection originated from the deepest map (i.e., the one maximizing the local significance) is selected and retained, while duplicate entries from shallower pointings are discarded.
To prevent spurious edge detections and artifacts from entering the matching pipeline, circular masks are preemptively applied in the noisy areas sorrounding the few very bright sources.
Fluxes are corrected for the PB attenuation. The profile of the PB correction is initially modeled and extrapolated using a Gaussian function and then empirically refined through successive iterations by tracking the flux stability of matching sources as a function of their radial distance from the respective pointing centers. 

Because nominal uncertainties from \texttt{SExtractor} often underestimate the true noise floor in the presence of correlated noise, we implemented an empirical self-calibration scheme. The pipeline monitors a subset of sources independently detected across overlapping maps and compares the standard deviation of their measured fluxes with their \texttt{SExtractor}-estimated errors. Based on the average ratio between this observational scatter and the nominal uncertainties, a global scaling factor is derived and applied to all flux errors in the catalog.

\section{The ``extended'' samples north and south}
\label{SECT:AppendixB}

In this section, we present the sources belonging to the northern and southern ``extended'' samples, each comprising the sources included in the corresponding ``core'' sample and the additional selected sources.

\begin{figure*}[!ht]
\centering
\includegraphics[width=17cm]{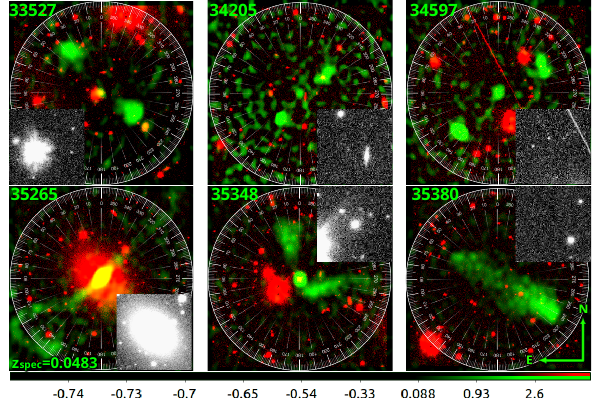}
\caption{Composite of the individual $170'' \times 170''$ radio stamps (in green) of some of the northern region sources excluded from the ``core'' sample and included in the ``extended'' sample. All visual conventions are identical to those defined in Figure~\ref{img:stacked_north_sample_all}. These sources are omitted primarily due to their smaller angular size (IDs 33527, 34205, 34597) or because of their unclear shape (IDs 35265, 35348, 35380). In this group, only IDs 33527 and 35380 show a significant misalignment relative to the stacked sample.  A clear VST detection is visible for IDs 35265 and 35348, while for ID 33527 the optical counterpart shows an offset with the radio counterpart, indicating a possible geometrical only (i.e., un-physical) association. While the optical counterpart does not show a clear shape for 33527 and 35348, the major axes is approximately orthogonal to the orientation of the jets in 35265. Source ID 35380 does not show a clear radio central emission where we could search for counterparts.}

\label{img:not_stacked_north_sample_all}
\end{figure*}

\begin{figure}[!ht]
\centering
\resizebox{\hsize}{!}{\includegraphics{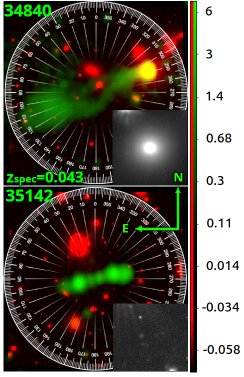}}
\caption{Composite of the individual $170'' \times 170''$ radio stamps  (in green) of two of the northern region sources excluded from the ``core'' sample and included in the ``extended'' sample. All visual conventions are identical to those defined in Figure~\ref{img:stacked_north_sample_all}. ID 34840 was excluded due to its asymmetrical morphology, while ID 35142 was omitted due to its smaller angular size. Both the two sources are aligned with the jets included in the stacked sample. A clear otical detection is visible only for 34840, with the jet roughly aligned with the semiminor axis.}
\label{img:not_stacked_north_sample_B}
\end{figure}

\begin{figure*}[!ht]
\centering
\includegraphics[width=15.5cm]{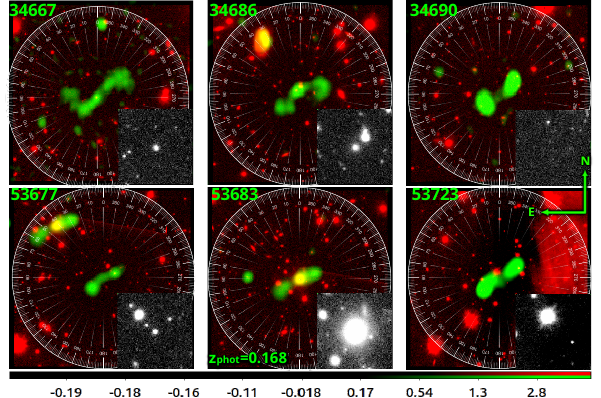}
\caption{Individual $170'' \times 170''$ radio stamps (in green) of the first half of the southern sample sources excluded from the ``core'' sample and included in the ``extended'' sample. All visual conventions are identical to those defined in Figure~\ref{img:stacked_north_sample_all}. All jets in this figure are aligned with the DM filament. An optical detection is visible for four out of six sources (34667, 34686, 53677, and 53683). The optical counterpart is clearly elongated only for ID 53683 (with the optical major axis roughly aligned with the DM filament.}
\label{img:NOT_stacked_south_sample_A}
\end{figure*}

\begin{figure*}[!ht]
\sidecaption
\includegraphics[width=10.5cm]{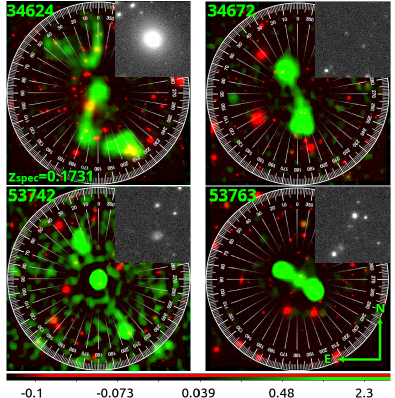}
\caption{Composite of the individual $170'' \times 170''$  radio stamps (in green) of the second half of the sample of southern region sources excluded from the ``core'' sample and included in the ``extended'' sample. All visual conventions are identical to those defined in Figure~\ref{img:stacked_north_sample_all}. In this sample we include sources with disturbed or asymmetric morphologies (IDs: 34624, 34672), or completely out of alignment with the filament. ID 53742 met all the ``core'' selection criteria but was not stacked due to its misalignment with the other selected sources. Only for ID 34624, a possible optical counterpart ($z=0.1731$) is identified at $\alpha_{J2000} = 13^{\text{h}} 28^{\text{m}} 27.3799^{\text{s}}, \delta_{J2000} = -32^{\circ} 40' 03.781''$.}
\label{img:NOT_stacked_south_sample_B}
\end{figure*}

Table~\ref{tbl:redshifts} summarizes the main properties of all sources analyzed in this work. In addition to positions, redshifts, and total fluxes, the table reports several characteristics illustrated in Figures~\ref{img:stacked_north_sample_all}, \ref{img:stacked_south_sample_all}, \ref{img:not_stacked_north_sample_all}, \ref{img:not_stacked_north_sample_B}, \ref{img:NOT_stacked_south_sample_A}, and \ref{img:NOT_stacked_south_sample_B}. Specifically, we report whether each source belongs to a ``core'' sample, its jet $PA$, and its apparent size. The presence of a VST counterpart is also indicated.

For detailed descriptions of the sources in the ``core'' samples, see Sections~\ref{sec:EXT_NORTH_SAMPLE} and \ref{sec:EXT_SOUTH_SAMPLE}. Here below, we describe the remaining sources in the ``extended'' samples. The additional sources in the northern sample are shown in Figures~\ref{img:not_stacked_north_sample_all} and \ref{img:not_stacked_north_sample_B}, while those in the southern sample are shown in Figures~\ref{img:NOT_stacked_south_sample_A} and \ref{img:NOT_stacked_south_sample_B}.

The complex source ID 35265 is situated in the western sector of the northern region. The main direction of its jets appears to be coupled with the western portion of the X-ray emission, mirroring the alignment seen in the ``core'' sample. Its spectroscopic redshift ($z=0.0483$) confirms that this source is gravitationally bound to the foreground Shapley complex. 

The v-shaped source ID 35348 (specifically, a Wide-Angle Tail) is located in the eastern sector of the northern region. The southern jet of this source appears initially directed along the NE-SW direction, matching the behavior of the northern jet, before undergoing a sharp $90^{\circ}$ deflection, ultimately aligning with the orientation of the filament. This morphology may suggest a transition from a jet-dominated ballistic phase to a regime governed by the ram pressure exerted by the intra-filament medium.

Concerning the remaining sources in the ``extended'' sample north, ID 35380 does not meet the morphological requirements but fulfills the other selection criteria. It exhibits one of the two largest angular separations from the filament barycenter among all candidates in the northern region and exhibits a significant misalignment relative to the filament. ID 34205, located peripherally, was excluded from the ``core'' sample due to its small angular extension ($\sim 37''$), though its orientation remains consistent with the general alignment within $\sim 20^{\circ}$. ID 33527 is situated near the ``core'' sample and shows a significant misalignment relative to the filamentary structure (angular size: $53''$). Conversely, ID 34597 is located in close proximity to the aligned sample and exhibits an orientation coherent with the filament (angular size: $56''$). ID 34840 shows a unidirectional jet emission that is apparently aligned with the filamentary axis (angular size: $70''$). Finally, ID 35142 is located near the filament, displaying an orientation consistent with the filament itself (angular size: $39''$).

Figure~\ref{img:NOT_stacked_south_sample_A} illustrates more than half of the entire collection of the sources failing to meet the ``core'' selection criteria in the southern region (IDs: 34667, 34686, 34690, 53677, 53683, 53723). Their angular size ranges from $\sim 25''$ to $35''$. Despite their smaller size, these sources exhibit orientations consistent with the stacked sample, further supporting the prevalence of the large-scale alignment effect beyond our primary selection criteria.

The second half of the sources excluded from the stacking analysis is shown in Figure~\ref{img:NOT_stacked_south_sample_B}. This group includes sources with disturbed or asymmetric morphologies (IDs: 34624, 34672), as well as those showing significant misalignment with the filament (IDs: 34672, 53742, 53763). Source ID 53742 (size $\sim 55''$), despite belonging to the ``core'' sample, is shown in this group due to its complete misalignment with the filament.

\begin{table*}
  \small
\caption{\label{tbl:redshifts}Known redshifts and source properties.}
\centering
\begin{tabular}{lllllcccll}
\hline
\hline
   &    &     & Radio &   &       & Meeting   &              &          & Linear    \\
ID & RA & Dec & flux  & z &  Type & ``core'' & VST$^a$      & Jets $PA \pm a$ $^b$    & Size$^c$    \\
   &    &     & [mJy] &   &       & criteria  &              & &  [$^{\prime\prime}$]/[Kpc]    \\
\hline
\multicolumn {10}{c}{North -- stacked sample}\\
\hline
19874 & 13:27:44.7095 & -30:18:16.812 & $980.2 \pm 3.6$    & 0.20009                & spec & yes & yes & $110^{\circ}\pm10^{\circ}$ & 81 / 276 \\
35274 & 13:29:53.5100 & -30:56:22.165 & $149.64 \pm 0.19$  &                        &      & yes & yes & $105^{\circ}\pm10^{\circ}$ & 65 \\
35311 & 13:30:39.9175 & -30:47:05.640 & $132.78 \pm 0.30$  & 0.2574$^d$             & spec & yes & yes & $120^{\circ}\pm15^{\circ}$ & 77 / 317 \\
      &               &               &                    & 0.239$\pm0.04^e$       & phot &     &     &                       & --- / 301\\
35350 & 13:28:19.2672 & -30:37:36.627 & $19.84 \pm 0.17$   &                        &      & yes & ?   &  $115^{\circ}\pm10^{\circ}$ & 73 \\
35360 & 13:25:45.7785 & -30:32:33.820 & $20.64 \pm 0.21$   & 0.5599$^d$             & spec & yes & yes & $105^{\circ}\pm15^{\circ}$  & 60 / 400 \\
35364 & 13:26:40.7641 & -30:32:15.375 & $5.24 \pm 0.13$    &                        &      & yes & yes & $100^{\circ}\pm15^{\circ}$   & 62       \\
\hline
\multicolumn {10}{c}{North - not stacked sample}\\
\hline
33527 & 13:27:53.2776 & -30:42:23.007 & $4.808 \pm 0.080$ &                        &      & no  & no  & $35^{\circ}\pm10^{\circ}$  & 53 \\
34205 & 13:28:42.0199 & -30:09:43.554 & $7.21 \pm 0.29$   &                        &      & no  & no  & $140^{\circ}\pm10^{\circ}$ & 37 \\
34597 & 13:28:03.2215 & -30:24:51.143 & $7.50 \pm 0.13$   &                        &      & no  & no  & $135^{\circ}\pm10^{\circ}$ & 56 \\
35265 & 13:26:18.7343 & -30:57:37.765 & $7.248 \pm 0.059$ & 0.0483$\pm 0.0001 ^f$  & spec & no  & yes & $130^{\circ}\pm10^{\circ}$ & 125/122\\
35380 & 13:25:51.0255 & -30:23:31.588 & $26.50 \pm 0.43$  &                        &      & no  & no  & $60^{\circ}\pm15^{\circ}$ & 60 \\
34840 & 13:28:02.6474 & -31:45:20.539 & $35.06 \pm 0.11$  & 0.04289$\pm 0.001 ^e$  & spec & no  & yes & $110^{\circ}\pm10^{\circ}$& 70/61\\
35142 & 13:29:32.2814 & -31:16:55.954 & $27.172 \pm 0.052$&                        &      & no  & ?   & $105^{\circ}\pm10^{\circ}$& 39 \\
\multicolumn {10}{c}{North - NOT stacked - complex shape}\\
35348 & 13:29:16.7045 & -30:38:25.481 & $8.42 \pm 0.15$   &                        &      & no  & yes & v-shaped              & 64 \\ 
\hline
\multicolumn {10}{c}{South -- stacked sample}\\
\hline
53568 & 13:31:46.7974 & -32:53:39.744 & $12.08 \pm 0.41$  & 0.0497$\pm 0.0001 ^f$  & spec & yes & yes & $135^{\circ}\pm20^{\circ}$& 54/54\\
53600 & 13:30:52.8890 & -32:35:21.127 & $3.02 \pm 0.11$   &                        &      & yes & ? & $125^{\circ}\pm10^{\circ}$  & 47\\
53617 & 13:33:14.6758 & -32:29:26.892 & $69.95 \pm 0.14$  &                        &      & yes & yes & $130^{\circ}\pm10^{\circ}$& 80\\
53619 & 13:32:14.1649 & -32:28:59.545 & $31.543 \pm 0.095$&                        &      & yes & yes & $125^{\circ}\pm10^{\circ}$& 53\\
\hline
\multicolumn {10}{c}{South -- not stacked sample}\\
\hline
53683 & 13:33:14.1978 & -32:12:08.265 & $5.204 \pm 0.023$  & 0.204$^d$              & spec & no  & yes & $120^{\circ}\pm15^{\circ}$& 28/97\\
      &               &               &                    & 0.16837$^e$            & phot &     &     &                      & ---/83\\
34667 & 13:28:22.5335 & -32:16:14.009 & $2.700 \pm 0.086$  &                        &      & no  & yes & $135^{\circ}\pm10^{\circ}$& 35 \\
34672 & 13:28:38.8105 & -32:13:38.042 & $8.539 \pm 0.072$  &                        &      & no  & no  & $25^{\circ}\pm15^{\circ}$ & 40 \\
34686 & 13:28:05.6483 & -32:09:32.155 & $5.169 \pm 0.052$  &                        &      & no  & yes & $135^{\circ}\pm20^{\circ}$& 31 \\
34690 & 13:27:52.1416 & -32:08:41.762 & $11.727 \pm 0.050$ &                        &      & no  & no  & $125^{\circ}\pm15^{\circ}$& 30 \\
53677 & 13:33:10.0345 & -32:12:51.576 & $3.005 \pm 0.027$  &                        &      & no  & yes & $130^{\circ}\pm15^{\circ}$& 25 \\
53723 & 13:31:42.6521 & -32:06:36.251 & $140.05 \pm 0.14$  &                        &      & no  & no  & $130^{\circ}\pm20^{\circ}$& 30 \\
53742 & 13:30:55.8904 & -32:04:07.700 & $11.910 \pm 0.035$ &                        &      & yes & ?   & $25^{\circ}\pm10^{\circ}$ & 55 \\
53763 & 13:32:41.1459 & -32:01:54.870 & $29.504 \pm 0.036$ &                        &      & no  & ?   & $60^{\circ}\pm10^{\circ}$ & 30 \\
\hline
\multicolumn {10}{c}{South - NOT stacked - complex shape}\\
\hline
34624 & 13:28:27.3799 & -32:40:03.781 & $110.09 \pm 0.74$  & 0.1731$\pm 0.0002 ^f$  & spec & no  & ?$^g$ & unclear & 74/225 \\
\hline
\end{tabular}

\tablefoot{ \\
\tablefoottext{a}{VST detection (yes/no). The position (Ra, Dec) of the most probable VST counterpart is reported when clearly identified;}
\tablefoottext{b}{Position angles measured on the east jets. The reported error represents a ``Type B'' uncertainty \citep{JCGM1002008}. The corresponding standard uncertainty can be obtained as $\sigma=a/\sqrt{3}$, with $a$ representing the half-width of the uncertainty interval.;}
\tablefoottext{c}{The apparent size is approximated within a few arcseconds ($\sim 3-5^{\prime\prime}$). A standard $\Lambda CDM$ model is considered ($H_{0}\sim67.4$, $\Omega m \sim0.315$) for the linear size; }
\tablefoottext{d}{From DESI, \citep{Wen2024}, reported as ``z cluster'';}
\tablefoottext{e}{From GLADE+ \citep{Dalya2022}, VizieR catalog VII/291;}
\tablefoottext{f}{From \cite{Haines2018};}
\tablefoottext{g}{Possible VST counterpart at 13:28:28.0415 -32:40:19.003.}
}
\end{table*}

\section{Binomial approach: significance through Monte Carlo simulations}

\label{SECT:AppendixD}

In Table~\ref{tab:stats_final_MC}, we report the results of the MC Simulations ($5\times10^9$ iterations) used to numerically test the analytical probabilities computed in Section~\ref{sec:stats}, Equations~\ref{EQ:P_fix} and \ref{EQ:P_any} \footnote{The software used to this purpose can be found at \url{https://github.com/Ibaronch/Papers}.}

\begin{table*}[h]
\small
\caption{\label{tab:stats_final_MC} Binomial alignment significance for different scenarios from MC simulations ($5\times10^9$ iterations)$^a$.} 
\centering 
\begin{tabular}{|l|cc|cc|} 
\hline \hline 
Scenario &  MC simulation         & Significance     & MC simulation        & Significance \\ 
         &    $P$(fixed $\theta$) & (fixed $\theta$) & $P$(any $\theta$)    & (any $\theta$) \\ 
\hline 
Core north (A)      & $(1.87 \pm 0.02) \times 10^{-6}$ & $4.6248 \pm 0.0021\sigma$ & $(1.02 \pm 0.001) \times 10^{-4}$ & $3.7151 \pm 0.0004\sigma$ \\ 
Extended north (B)  & $(3.21 \pm 0.03) \times 10^{-6}$ & $4.5119 \pm 0.0017\sigma$ & $(1.45 \pm 0.002) \times 10^{-4}$ & $3.6249 \pm 0.0003\sigma$ \\ 
Core south (C)      & $(6.95 \pm 0.004) \times 10^{-4}$ & $3.1969 \pm 0.0002\sigma$ & $(3.19 \pm 0.001) \times 10^{-3}$ & $2.7276 \pm 0.0001\sigma$ \\ 
Extended south (D)  & $(6.34 \pm 0.36) \times 10^{-8}$ & $5.28 \pm 0.01\sigma$ & $(4.04 \pm 0.09) \times 10^{-7}$  & $4.9333 \pm 0.0043\sigma$ \\ 
\hline
\textbf{Global (joint)} $^b$    & $(1.99 \pm 0.11) \times 10^{-13}$ & $\mathbf{7.26 \pm 0.01\sigma}$ & $(5.86 \pm 0.13) \times 10^{-11}$ & $\mathbf{(6.443 \pm 0.004)\sigma}$ \\
\textbf{Global (combined)} $^c$ & $(3.00 \pm 0.28) \times 10^{-10}$ & $\mathbf{6.14 \pm 0.11\sigma}$ & $(2.64 \pm 0.23) \times 10^{-8}$  & $\mathbf{(5.442 \pm 0.015)\sigma}$ \\ 
\hline
\end{tabular} 
\tablefoot{\\
\tablefoottext{a}{Computed with the Python script MC\_sim6.py, that can be found at \url{https://github.com/Ibaronch/Papers}; } 
\tablefoottext{b}{Calculated as the product of probabilities from Scenarios B and D, described in Section~\ref{sec:stats}; }
\tablefoottext{c}{Calculated through MC simulations, considering all sources in scenarios B and D, described in Section~\ref{sec:stats}.}
}
 
\end{table*}

\FloatBarrier 
\clearpage

\end{appendix}
\end{document}